\documentclass[aps,twocolumn, superscriptaddress, english, nofootinbib, 10pt]{revtex4-2}

\usepackage{float}

\usepackage{amssymb, amsmath, bm, dcolumn, epsf, graphicx, latexsym, slashed, simplewick}
\usepackage[utf8]{inputenc}
\usepackage{subfigure}
\usepackage{comment}
\usepackage{graphicx}
\usepackage{hyperref}
\usepackage{enumerate}
\usepackage{xspace}
\usepackage{fontawesome}

\usepackage[usenames,dvipsnames]{xcolor}
\hypersetup{
    colorlinks = true,
    citecolor = {MidnightBlue},
    linkcolor = {BrickRed},
    urlcolor = {BrickRed}
}

\newcommand{\ba}{\begin{eqnarray}}
\newcommand{\ea}{\end{eqnarray}}
\newcommand{\planck}{{\sl Planck}\xspace}
\newcommand{\lMc}{\log_{10}M_c}
\newcommand{\nv}{\hat{\bf n}}
\newcommand{\nmt}{{\tt NaMaster}\xspace}
\newcommand{\hpx}{{\tt HEALPix}\xspace}
\newcommand{\xray}{X-ray\xspace}
\newcommand{\xrays}{X-rays\xspace}

\newcommand{\crate}{{\rm CR}}
\newcommand{\energ}{\varepsilon}
\newcommand{\energo}{\energ_o}
\newcommand{\energm}{\tilde{\energ}_o}
\newcommand{\energe}{\energ_e}

\newcommand{\al}[1]{{\textcolor{black}{{#1}}}}

\begin{document}

\title{$X+y$: insights on gas thermodynamics from the combination of \xray and thermal Sunyaev-Zel'dovich data cross-correlated with cosmic shear}

\author{Adrien La Posta}
\email{adrien.laposta@physics.ox.ac.uk}
\affiliation{Department of Physics, University of Oxford, Denys Wilkinson Building, Keble Road, Oxford OX1 3RH, United Kingdom}
\author{David Alonso}
\affiliation{Department of Physics, University of Oxford, Denys Wilkinson Building, Keble Road, Oxford OX1 3RH, United Kingdom}
\author{Nora Elisa Chisari}
\affiliation{Institute for Theoretical Physics, Utrecht University, Princetonplein 5, 3584 CC, Utrecht, The Netherlands}
\affiliation{Leiden Observatory, Leiden University, Niels Bohrweg 2,2333 CA, Leiden, The Netherlands}
\author{Tassia Ferreira}
\affiliation{Department of Physics, University of Oxford, Denys Wilkinson Building, Keble Road, Oxford OX1 3RH, United Kingdom}
\author{Carlos Garc\'ia-Garc\'ia}
\affiliation{Department of Physics, University of Oxford, Denys Wilkinson Building, Keble Road, Oxford OX1 3RH, United Kingdom}

\begin{abstract}
  We measure the cross-correlation between cosmic shear from the third-year release of the Dark Energy Survey, thermal Sunyaev-Zel'dovich (tSZ) maps from \planck, and \xray maps from ROSAT. We investigate the possibility of developing a physical model able to jointly describe both measurements, simultaneously constraining the spatial distribution and thermodynamic properties of hot gas. We find that a relatively simple model is able to describe both sets of measurements and to make reasonably accurate predictions for other observables (the tSZ auto-correlation, its cross-correlation with \xrays, and tomographic measurements of the bias-weighted mean gas pressure). We show, however, that contamination from \xray AGN, as well as the impact of non-thermal pressure support, must be incorporated in order to fully resolve tensions in parameter space between different data combinations. \al{Combining the tSZ and \xray cross-correlations with cosmic shear} we obtain simultaneous constraints on the mass scale at which half of the gas content has been expelled from the halo, $\lMc=14.83^{+0.16}_{-0.23}$, on the polytropic index of the gas, $\Gamma=1.144^{+0.016}_{-0.013}$, and on the ratio of the central gas temperature to the virial temperature $\alpha_T=1.30^{+0.15}_{-0.28}$, \al{marginalizing over AGN contributions to the signal.} 
\end{abstract}

\maketitle

\section{Introduction}
  Baryons make up approximately $5\%$ of the total energy density of the Universe today \citep{1807.06209}. The majority of this contribution ($\sim80-90\%$) is in the form of ionised intergalactic gas, in both hot and warm phases \citep{astro-ph/0406095,astro-ph/0601008,0706.1787,1112.2706,2211.04058}. Despite their small contribution to the total energy budget, and the fact that baryons are governed by known fundamental physical processes, our relatively poor understanding of the distribution and thermodynamic properties of cosmic gas is currently one of the main roadblocks on the way towards precision cosmology from Stage-IV experiments. This is so in the context of weak lensing (due to the baryonic suppression of the matter power spectrum \citep{1105.1075,1810.08629,1905.06082}), Cosmic Microwave Background (CMB) secondary anisotropies (e.g. uncertainties in cluster mass-observable relations \citep{1812.01679,1904.07887,2010.07797} and degeneracies between optical depth and growth \citep{1604.01382,1607.02442,2101.08373,2305.00992}), \xray cluster science (e.g. hydrostatic mass biases \citep{1308.6589,2211.08372}), among others.

  The reason for this poor understanding is the high complexity of the physical systems and processes that govern gas on astrophysical scales, and the strong impact of phenomena taking place at very small scales on the large-scale distribution of gas. These processes include radiative gas cooling, gravitationally-driven virialisation, and non-gravitational heating from stars and active galactic nuclei (AGN). In particular, AGN feedback effects are the dominant source of uncertainty in the level of baryonic suppression of the power spectrum on small scales \citep{2206.11794,2206.08591,2303.05537,2403.13794,2404.06098}, and thus constitute an attractive explanation for the so-called ``$S_8$ tension'' between late-time weak lensing data and early-time CMB measurements \citep{2007.15632,2105.12108,2105.13549}. Developing data-driven methods to improve our understanding of these processes is therefore of paramount importance.

  Fortunately, the fast growth in the abundance and quality of wide-area multi-wavelength observations in the last few years has made available a wide array of probes that are sensitive to complementary physical properties of the cosmic gas. Perhaps the most prominent of these, in the context of cosmology, are \xray maps \citep{Voges:1993,2401.17274}, and measurements of the thermal and kinematic Sunyaev-Zel'dovich effects (tSZ and kSZ, respectively \citep{1502.01596,2307.01258}). In addition to these, near-future observations will enable novel tracers of gas, such as patchy screening \citep{0812.1566,2401.13033}, dispersion measure statistics from fast radio bursts \citep{1901.02418,2201.04142}, relativistic SZ \citep{1809.09666,2110.13932}, and indirect probes of feedback, such as multi-wavelength constraints on star formation \citep{2209.05472,2310.10848,2404.07252}. In this work, we will focus on the combination of tSZ and \xray data as a potential probe of gas properties. In particular, we will study the cross-correlations of these two gas tracers with weak gravitational lensing data in the optical (i.e. cosmic shear).
  
  Cross-correlations between shear and tSZ have been exploited in the literature as a probe of baryonic effects, and for their potential to improve cosmological constraints from weak lensing data due to the complementary dependence on cosmological parameters \citep{1310.5721,1910.07526,2108.01600,2108.01601,2109.04458}. A drawback of this approach is the degeneracy between gas density and temperature, since the SZ Compton-$y$ parameter is sensitive to the line of sight integral of the thermal gas pressure $P_{\rm th}\propto \rho_{\rm gas}\,T_{\rm gas}$, where $\rho_{\rm gas}$ and $T_{\rm gas}$ are the density and temperature of the gas. Thus, in the absence of additional information, a model able to connect the distribution of gas and its thermal state unambiguously is required to fully exploit this cross-correlation \citep{2005.00009,2406.01672}.

  \xray data has also been exploited in the context of baryonic physics, especially through the measurement of bound gas fractions from the observation of individual galaxy clusters \citep{1712.05463,1911.08494,2110.02228} and, more recently, including derived constraints on the electron density profile \citep{2309.02920}. The cross-correlation of \xray maps with cosmic shear data itself was first presented in \cite{2309.11129}. Cross-correlations between \xray maps and other large-scale structure tracers have also been used in the literature for similar purposes \cite{astro-ph/0302067,1505.03658,1701.09016,1701.09018,1711.10774,1902.08268,1909.02179,2204.13105}. Potential advantages of the cross-correlation approach are the absence of selection effects, and the ability to incorporate the measurements in a joint cosmological analysis with cosmic shear and galaxy clustering data, accounting for all data correlations in a completely consistent manner. Since the \xray emissivity $j_\varepsilon$ has a complementary dependence on gas density and temperature ($j_\varepsilon\propto \rho_{\rm gas}^2\Lambda_c(T_{\rm gas})$, where $\Lambda_c$ is the \xray cooling function \citep{astro-ph/0512549}), a combination of \xray and tSZ data could be a powerful way to break the degeneracy between gas density and thermodynamics. Furthermore, the strong dependence on gas density ($\propto\rho_{\rm gas}^2$) makes \xray observations more sensitive to the inner regions of dark matter haloes, whereas tSZ data is also sensitive to the outskirts, adding to the complementarity of both probes. Both probes are also sensitive to different sources of contamination, such as the Cosmic Infrared Background (CIB) and Galactic dust for tSZ \citep{mccarthy2024}, and emission from unresolved AGN in the case of \xray data\footnote{Some of these AGN may exhibit radio emission and contaminate tSZ measurements at low frequencies, although this contribution is usually subdominant to dust.} \citep{1503.01120,1608.05184,2301.01388}.

  The main aim of this paper is to explore the possibility of describing the \xray and tSZ cross-correlations within a single hydrodynamical model of gas and to constrain it from data. Of particular interest is the performance of such a model in the presence of non-negligible sources of observational and theoretical uncertainty that affect each probe differently, such as contamination from \xray AGN or the impact of non-thermal pressure support. In this sense, the use of cross-correlations with cosmic shear makes this a simpler task since, on small scales, gas properties need only be related to the matter density, rather than more physically complex quantities, such as galaxy abundance, which may require additional complexity in the model \citep{koukoufilippas2019}.

  This paper is structured as follows. Section \ref{sec:meth} presents the hydrodynamical model used to describe the measured cross-correlations, as well as the methods used to make these measurements. The datasets used in our analysis are described in Section \ref{sec:data}. Section \ref{sec:res} presents our results, including the model constraints, the ability of the best-fit model to describe other gas observables, and the impact of observational and theoretical uncertainties. We then conclude in Section \ref{sec:conc}. We use natural units throughout, with $c=1$, unless otherwise stated.

\section{Methods}\label{sec:meth}
\subsection{Hydrodynamical halo model for gas physics}\label{ssec:meth.halomodel}
  We use the halo model to describe tSZ and \xray signal from warm/hot diffuse gas. Specifically, our fiducial model is that used in \cite{2309.11129}, which we describe in more detail below. The model itself is heavily inspired on the hydrodynamical halo model of \cite{2005.00009}, as well as the so-called ``Baryon Correction Model'' (BCM) of \cite{1510.06034}.
  
  We model the total matter density as a sum of different contributions from cold dark matter (\textbf{CDM}), stars ($\mathbf{\star}$), bound (\textbf{b}) and ejected gas (\textbf{e}). The density profile for each component is characterised by a mass-dependent mass fraction $f_x(M)$, and a scale-dependent function $g_x(r|M)$, where $r$ is the comoving distance to the halo center. The physical mass density in component $x$ is thus given by
  \begin{equation}
    \rho_x(r|M)=\frac{M}{a^3}\,f_x(M)\,g_x(r|M),
  \end{equation}
  where $a$ is the scale factor, and the scale-dependent function is normalised to
  \begin{equation}\label{eq:gnorm}
    4\pi\,\int_0^{\infty} dr\,r^2g_x(r|M)=1.
  \end{equation}

  The next sections present the models used for $f_x(M)$ and $g_x(r|M)$ for the different components. We also require a model for the thermal state of the gas, which we parametrise in terms of the gas pressure. This is described below for the bound and ejected gas components.

  \subsubsection{Cold dark matter}\label{sssec:meth.halomodel.cdm}
    The fractional dark matter abundance is fixed to the global cosmic abundance $f_\textbf{CDM}(M)=\Omega_c/\Omega_m$, where $\Omega_c$ and $\Omega_m$ are the fractional energy densities in cold dark matter and total non-relativistic matter, respectively.

    We parametrise the scale dependence using a truncated Navarro-Frenk-White profile of the form
    \begin{equation}
      g_\textbf{CDM}(r|M)=\frac{1}{V_\textbf{CDM}}\frac{\Theta(r<r_\Delta)}{\frac{r}{r_s}\left(1+\frac{r}{r_s}\right)^2}.
    \end{equation}
    Here, $r_\Delta$ is the virial radius, defined as the radius enclosing a total mass that is $\Delta=200$ times larger than the mean critical density of the Universe. The scale radius $r_s$ is related to the virial radius by the concentration parameter $c_\Delta(M)$ via $r_\Delta=c_\Delta\,r_s$. $\Theta$ is a Heavyside function, and the proportionality factor $V_\textbf{CDM}$ enforces the normalisation in Eq. \ref{eq:gnorm}. For the NFW profile, this is given by
    \begin{equation}
      V_\textbf{CDM}=4\pi r_s^3\left[\log(1+c_\Delta)-\frac{c_\Delta}{1+c_\Delta}\right].
    \end{equation}
    The Fourier transform of this profile, which is needed to construct halo model power spectra (see Section \ref{sssec:meth.halomodel.halomodel}), can be calculated analytically (see e.g. \cite{koukoufilippas2019}).    
    
  \subsubsection{Stars}\label{sssec:meth.halomodel.star}
    The stellar mass fraction is modelled as in \cite{1406.5013}
    \begin{equation}
      f_*(M) = A_*\,\exp\left[-\frac{1}{2}\left(\frac{\log_{10}(M/M_*)}{\sigma_*}\right)^2\right],
    \end{equation}
    where $M_*=10^{12.5}M_\odot$, $\sigma_*=1.2$, and $A_*=0.03$ \citep{1205.5807,1401.7329}. Furthermore, we assume that stars contribute only at very small distances from the central galaxy, and thus approximate the scale-dependent profile as a Dirac delta function at $r=0$. The corresponding Fourier-space profile is therefore simply a constant:
    \begin{equation}
      \rho_*(k|M)=\frac{M}{a^3}f_*(M).
    \end{equation}

  \subsubsection{Bound gas}\label{sssec:meth.halomodel.bound}
    The diffuse gas component is divided into a ``bound'' contribution, corresponding to virialised gas that has not been expelled from the halo, and a low-density ``ejected'' contribution, expelled outside of the halo virial radius by AGN-driven outflows.

    The mass-dependence of the bound fraction is modeled as
    \begin{equation}
      f_\textbf{b}(M)=\frac{\Omega_b/\Omega_m}{1+(M_c/M)^\beta},
    \end{equation}
    where $\Omega_b$ is the cosmic baryon abundance, $M_c$ is the mass scale at which half of the gas has been expelled, and the slope $\beta$ determines the steepness of the curve. Effectively, $M_c$ determines the amount of bound gas at sufficiently high density and temperature to contribute significantly to both tSZ and \xray measurements, and hence plays an important role in governing the amplitude of both cross-correlations. We will refer to $M_c$ as the ``halfway mass'' in what follows. We find that our measurements are relatively insensitive to the value of the mass slope $\beta$, and hence we fix it to $\beta=0.6$ \citep{1510.06034}.

    The scale dependence of the bound gas is given by the Komatsu-Seljak (KS) \citep{astro-ph/0205468} profile. This is determined under the assumption of hydrostatic equilibrium balancing gas pressure and gravity:
    \begin{equation}
      \frac{dP}{dr}=-\frac{GM(<r)\rho_\textbf{b}(r)}{r^2},
    \end{equation}
    where $M(<r)$ is the total mass enclosed within a radius $r$. Assuming a polytropic equation of state $P\propto\rho_\textbf{b}^\Gamma$ with a polytropic index $\Gamma$, and an NFW profile to calculate $M(<r)$, the equation above can be solved for $\rho_\textbf{b}$ to yield a relatively simple scale dependence:
    \begin{equation}
      g_\textbf{b}(r|M)=\frac{1}{V_\textbf{b}}\left[\frac{\log(1+r/r_s)}{r/r_s}\right]^{\frac{1}{\Gamma-1}}.
    \end{equation}
    The normalisation-enforcing prefactor in this case is
    \begin{equation}
      V_\textbf{b}\equiv4\pi r_s^3\,I\left(\frac{1}{\Gamma-1},0\right),
    \end{equation}
    where we have defined the integral
    \begin{equation}
      I(\gamma,q)\equiv\,\int_0^\infty dx\,x^2\left(\frac{\log(1+x)}{x}\right)^\gamma j_0(qx),
    \end{equation}
    with $j_0(x)\equiv\sin(x)/x$.

    The Fourier transform of this profile must be calculated numerically for general values of $\Gamma$. Fortunately, it can be expressed as a function of $\gamma\equiv1/(\Gamma-1)$ and the combination $q\equiv kr_s$ alone as:
    \begin{equation}
      g_\textbf{b}(k|M)=\frac{I(\gamma,q)}{I(\gamma,0)}.
    \end{equation}
    To speed up the evaluation of theoretical predictions, we precalculate $I(\gamma,q)$ and interpolate it as a function of $\gamma$ and $q$.

    The two parameters governing the bound gas density are therefore $M_c$, determining the overall abundance of virialised gas, and $\Gamma$, describing the scale dependence of its distribution.

  \subsubsection{Ejected gas}\label{sssec:meth.halomodel.ejected}
    Once the mass fractions for the stellar and bound components are known, the ejected mass fraction is simply
    \begin{equation}
      f_\textbf{e}(M)=1-f_\textbf{CDM}(M)-f_*(M)-f_\textbf{b}(M).
    \end{equation}
    For the scale dependence, we use the Gaussian model of \cite{1510.06034} (motivated under the assumption of a Maxwellian distribution for the velocities that cause the gas to be ejected):
    \begin{equation}
      g_\textbf{e}(r|M)=\frac{\exp[-r^2/(2r_e^2)]}{(2\pi r_e^2)^{3/2}},
    \end{equation}
    the characteristic scale $r_e$ is connected with the typical distance travelled by a gas particle travelling at the escape velocity. As in \cite{1510.06034}, this is parametrised as:
    \begin{align}
      r_e=0.375\eta_b\sqrt{\Delta}\,r_\Delta,
    \end{align}
    where $\Delta=200$ in our case, and $\eta_b$ quantifies the distance to which gas is expelled by AGN outflows. The Fourier transform of this profile is analytical $g_\textbf{e}(k|M)=\exp[-(kr_e)^2/2]$. We find that our data is insensitive to the value of $\eta_b$, since the ejected gas is too diffuse to produce appreciable \xray emission, and too cold to contribute significantly to the tSZ sigal. We thus fix this parameter to $\eta_b=0.5$ in our analysis consistent with the range of values presented in~\cite{1510.06034}.

  \subsubsection{Temperature, pressure, and number density}\label{sssec:meth.halomodel.TPn}
    The observables under study in this work are mostly sensitive to the number densities and temperature of different particle species in the bound gas.

    The number density of particle type $q$ is related to the gas mass density $\rho_\textbf{g}=\rho_\textbf{b}+\rho_\textbf{e}$ through its mean molecular weight $\mu_q$ via $n_q=\rho_\textbf{g}/(\mu_q\,m_p)$, where $m_p$ is the proton mass. Assuming a fully ionised gas made out of free electrons, hydrogen, and helium nuclei, the relevant $\mu_q$s are:
    \begin{equation}
      \mu_{\rm H}=\frac{1}{X_{\rm H}},\hspace{12pt}\mu_e=\frac{2}{1+X_{\rm H}},\hspace{12pt}\mu_T=\frac{4}{3+5X_{\rm H}},
    \end{equation}
    where $X_{\rm H}$ is the hydrogen mass fraction, and $\mu_T$ is the mean molecular weight of all particle species combined.

    As described in Section \ref{sssec:meth.halomodel.bound}, the bound gas is assumed to have a polytropic equation of state, relating its pressure to its density $P\propto \rho^\Gamma$. Assuming the bulk of this pressure to be thermal pressure, the temperature could then be calculated from the relation $P_{\rm th}=k_B\,n\,T\propto\rho\,T$. In this case $T\propto\rho^{\Gamma-1}$, and the scale dependence of the temperature is univocally determined to be $T(r)\propto\Phi(r)\propto\log(1+r/r_s)/(r/r_s)$, where $\Phi(r)$ is the gravitational potential of the NFW profile. In order to allow for the presence of non-thermal pressure support, we instead parametrise the temperature of the bound component as
    \begin{equation}
      T_\mathbf{b}(r) = T_c\left[ \frac{\mathrm{log}(1+r/r_s)}{r/r_s} \right]^{\gamma_T},
    \end{equation}
    where $\gamma_T \neq 1$ quantifies the presence of non-thermal pressure. The central temperature $T_c$ should be of the order of the virial temperature of the gas, and therefore we parametrise it, as in \cite{2005.00009}, in terms of a free, order 1, parameter $\alpha_T$ as
    \begin{equation}
      k_B T_c = \alpha_T \frac{2}{3}\frac{GM\,m_p\mu}{ar_\Delta},
    \end{equation}
    where $\mu_T=0.61$, as in \cite{2005.00009}, is the mean molecular weight, and $a$ is the scale factor.

    Note that, within this model, the thermal pressure fraction is given by
    \begin{equation}\label{eq:nt_pres_definition}
      \frac{P_{\rm th}(r)}{P(r)}=\left[\frac{\log(1+r/r_s)}{r/r_s}\right]^{\gamma_T-1}.
    \end{equation}
    This scale dependence is similar to that found in \cite{2002.01934}, out to the virial radius $r=r_\Delta$, for $\gamma_T\sim1.5$, with the advantage of retaining the same simple functional form for the thermal pressure profile used for all calculations involving the bound gas component ($P_{\rm th}\propto[\log(1+x)/x]^{\gamma'}$ with $\gamma'=1/(\Gamma-1)+\gamma_T$). Our fiducial analysis will assume no non-thermal pressure support (i.e. we will fix $\gamma_T=1$), and we will study the effect of allowing for non-thermal pressure in Section~\ref{sssec:res.params.nonth}.

    Finally, we fix the temperature of the ejected gas to that of the warm-hot intergalactic medium $T_\mathbf{e}=10^{6.5}\,{\rm K}$ \citep{astro-ph/9806281,1310.5721}.

  \subsubsection{Halo model ingredients}\label{sssec:meth.halomodel.halomodel}
    Within the halo model \citep{astro-ph/0001493,astro-ph/0005010,astro-ph/0206508}, the power spectrum between two fields $U({\bf x})$ and $V({\bf x})$ is a combination of two-halo and one-halo correlations:
    \begin{equation}
      P_{UV}(k)= \langle bU\rangle\langle bV\rangle P_{\rm lin}(k)+P^{1h}_{UV}(k),
    \end{equation}
    where
    \begin{align}\label{eq:bU}
      &\langle bU\rangle\equiv\int dM\,n(M)\,b_h(M)\,\langle U(k|M)\rangle,\\\label{eq:pk1h}
      &P^{1h}_{UV}(k)\equiv\int dM\,n(M)\,\langle U(k|M) V(k|M)\rangle.
    \end{align}
    Here, $P_{\rm lin}(k)$ is the linear matter power spectrum, $n(M)$ and $b_h(M)$ are the halo mass function and the linear halo bias, $U(k|M)$ is the Fourier-space profile of quantity $U$ around halos of mass $M$, and the angle brackets imply averaging over halos of the same mass. The models used for the halo profiles of the different quantities used in this analysis have been described in the previous sections. In our theory predictions, we will use the halo mass function of \cite{0803.2706}, the halo bias parametrisation of \cite{1001.3162}, and the concentration-mass relation of \cite{0804.2486}. We use a spherical overdensity halo mass definition, with an overdensity parameter $\Delta=200$, defined with respect to the critical density. 

\subsection{\xray, tSZ, and cosmic shear}\label{ssec:meth.xys}
  This section presents the models used to construct theoretical predictions for our measured cross-correlations involving \xray count-rate maps, tSZ Compton-$y$ maps, and cosmic shear data.

  \subsubsection{Angular power spectra}\label{sssec:meth.xys.cls}
    Let $u(\nv)$ be a field defined on the celestial sphere, and related to a three-dimensional quantity $U({\bf x},z)$ through a line-of-sight projection of the form
    \begin{equation}\label{eq:projfield}
      u(\nv)=\int d\chi\,W_u(\chi)\,U(\chi\nv,z(\chi)),
    \end{equation}
    where $\chi$ is the comoving distance, $z(\chi)$ is the corresponding redshift in the lightcone, and $W_u(\chi)$ is the \emph{radial kernel} associated with $u$.
    
    The angular power spectrum of two such quantities, $u$ and $v$, $C^{uv}_\ell$, is related to the power spectrum of their three-dimensional counterparts $P_{UV}(k,z)$ via
    \begin{equation}
      C_\ell^{uv}=\int \frac{d\chi}{\chi^2}\,W_u(\chi)\,W_v(\chi)\,P_{UV}\left(k\equiv\frac{\ell+1/2}{\chi},z(\chi)\right).
    \end{equation}
    This equation is valid in the Limber approximation \cite{1953ApJ...117..134L}, which is sufficiently accurate for the broad radial kernels of the quantities explored in this work.

    The angular power spectrum is therefore determined by the radial kernels of the quantities involved, and the power spectrum of their three-dimensional counterparts $P_{UV}$. The model used to estimate $P_{UV}$ was described in Section \ref{ssec:meth.halomodel}. The following sections describe the kernels and 3D quantities associated with the three fields studied in this work.

  \subsubsection{Compton-$y$ tSZ maps}\label{sssec:meth.xys.y}
    CMB photons are inverse-Compton scattered by thermal free electrons in the intergalactic medium (IGM), particularly in galaxy clusters and groups, through the so-called thermal Sunyaev-Zel'dovich effect \cite{1972CoASP...4..173S}. The SZ effect modifies the CMB spectrum in a universal way, and thus multi-frequency observations can be used to separate the secondary anisotropies induced by this scattering. These are represented by the Compton-$y$ parameter, given by a line of sight integral of the electron thermal pressure:
    \begin{equation}
      y(\nv)=\int \frac{d\chi}{1+z}\,\frac{\sigma_T}{m_ec^2}P_e^{\rm th}(\chi\nv,z(\chi)).
    \end{equation}
    where $\sigma_T$ is the Thomson scattering cross section. Thus, the three-dimensional quantity associated with tSZ observations is the thermal electron pressure, and the radial kernel is
    \begin{equation}
      W_y(\chi)\equiv\frac{\sigma_T}{m_ec^2(1+z)}.
    \end{equation}
    The thermal electron pressure profile is computed separately for the bound and ejected gas components combining their density and temperature profiles, using $P_e^{\rm th}=k_Bn_eT_e$. The total pressure is then the sum of the pressures of both components.

  \subsubsection{\xray count rate maps}\label{sssec:meth.xys.x}
    We use \xray data in the form of count rate maps, i.e. the number of photons observed per unit time and solid area in a given energy band:
    \begin{equation}
      \crate(\nv)\equiv\frac{dN_o}{dt_od\Omega_o},
    \end{equation}
    where the subscript $_o$ denotes quantities measured in the observer's frame. In this context, it is important to distinguish between \emph{observer-frame energy} $\energo$ (i.e. the true energy of the photon as it hits the detector), and the \emph{measured energy} $\energm$ (i.e. the energy measured by the instrument). The relationship between both quantities is usually quantified in terms of the so-called energy redistribution matrix, the probability of measuring a given $\energm$ if the true energy is $\energo$:
    \begin{equation}
      {\cal M}(\energm|\energo)\equiv \frac{dp(\energm|\energo)}{d\energm}.
    \end{equation}
    In addition to this, the effective area of \xray detectors depends on the incident photon energy $A(\energo)$. Thus, the observed count rate is related to the photon specific intensity (i.e. number of photons per unit time, solid angle, area, and energy interval) in the observer's frame via
    \begin{equation}
      \crate(\nv)=\int d\energo\,\phi(\energo)\,A(\energo)\,\frac{dN_o}{d\energo\,dt_o\,dA_o\,d\Omega_o},
    \end{equation}
    where we have defined the instrument bandpass $\phi(\energo)$ as
    \begin{equation}
      \phi(\energo)\equiv\int_{\tilde{\energ}_{o,{\rm min}}}^{\tilde{\energ}_{o,{\rm max}}} d\energm\,{\cal M}(\energm|\energo),
    \end{equation}
    with $(\tilde{\energ}_{o,{\rm min}},\tilde{\energ}_{o,{\rm max}})$ the edges of the observed band.

    Now, the contribution to the specific intensity from sources at a comoving distance interval $(\chi,\chi+d\chi)$ is
    \begin{equation}
      d\left[\frac{dN_o}{d\energo dt_o dA_o d\Omega_o}\right]=\frac{d\chi}{4\pi(1+z)^3}j_\energ(\chi\nv,z,\energe),
    \end{equation}
    where $j_\energ\equiv dN_e/(d\energe dt_e dV_e)$ is the emissivity (number of photons emitted per unit energy, time, and physical volume), and the subscript $_e$ denotes quantities in the emitter's reference frame. Thus, the count rate map is
    \begin{align}\nonumber
      \crate(\nv)=\int \frac{d\chi}{4\pi(1+z)^3}\,&\int d\energo\,\phi(\energo)A(\energo)\\\label{eq:crgen}
      &\hspace{12pt}\times j_\energ(\chi\nv,z,\energo(1+z)).
    \end{align}
    The specific form of $j_\energ$ depends on the type of source. Here we will consider two cases:
    \begin{itemize}
      \item {\bf Diffuse gas.} Our main case of interest is \xray emission by hot IGM gas. This is dominated by collisional processes, prominently bremsstrahlung and collisionally-driven line emission \citep{astro-ph/0512549}. In this case, the emissivity is proportional to the square number density of gas particles and to the cooling function $\Lambda_c(\energ,T,Z)$, depending on the plasma temperature and metallicity. Specifically:
      \begin{equation}
        j_\energ({\bf x},\energ)=n_e({\bf x})n_{\rm H}({\bf x})\,\Lambda_c(\energ,T({\bf x}),Z({\bf x})).
      \end{equation}
      Inserting this in Eq. \ref{eq:crgen}, the count rate map for diffuse gas emission is
      \begin{equation}\label{eq:crgas}
        \crate_{\rm gas}(\nv)=\int\frac{d\chi}{4\pi(1+z)^3}\left[n_e\,n_{\rm H}\,J(T,Z,z)\right]_{(\chi\nv,z)},
      \end{equation}
      where we have defined
      \begin{equation}
        J(T,Z,z)\equiv\int d\energo\,\phi(\energo)\,A(\energo)\,\Lambda_c((1+z)\energo,T,Z).
      \end{equation}
      Thus, the radial kernel for \xray emission is simply
      \begin{equation}
        W_X(\chi)\equiv\frac{1}{4\pi(1+z)^3},
      \end{equation}
      and its associated three-dimensional quantity is the combination $n_e\,n_{\rm H}J(T,Z,z)$. We assume a metallicity $Z=0.3Z_\odot$, where $Z_\odot$ is the solar metallicity \citep{1811.01967}. We compute the cooling function using the Astrophysics Plasma Emission Code ({\tt APEC} \cite{astro-ph/0106478}) as implemented in {\tt PYATOMDB} \cite{1207.0576}.
      \item {\bf AGN.} A significant fraction of the diffuse \xray background is emission from unresolved non-thermal point sources, specifically AGN \citep{astro-ph/0501058,1311.1254,1503.01120,1901.10866}. Our analysis will therefore quantify their contribution to the measured \xray cross-correlations in different ways. The specifics of the model used here to estimate the contribution from unresolved AGN are described in Appendix \ref{app:agn}. In short, the contribution from AGN to the anisotropies in the count-rate map is given by
      \begin{equation}\label{eq:crAGN}
        \crate_{\rm AGN}(\nv)=\int d\chi\left\langle\frac{A}{\energ}\right\rangle_z\,\bar{\rho}_L(z)\,\delta_s(\chi\nv,z),
      \end{equation}
      where $\langle A/\energ\rangle_z$ is the value of the detector area over photon energy weighted by the source spectrum and averaged over the energy band (see Eq. \ref{eq:mean_Aoe}), $\bar{\rho}_L(z)$ is the background \xray luminosity density due to AGN (see Eq. \ref{eq:rhoL}), and $\delta_s$ is the overdensity of AGN (see Appendix \ref{app:agn}).
    \end{itemize}

  \subsubsection{Cosmic shear}\label{sssec:meth.xys.s}
    \begin{figure}
        \centering
        \includegraphics[width=\linewidth]{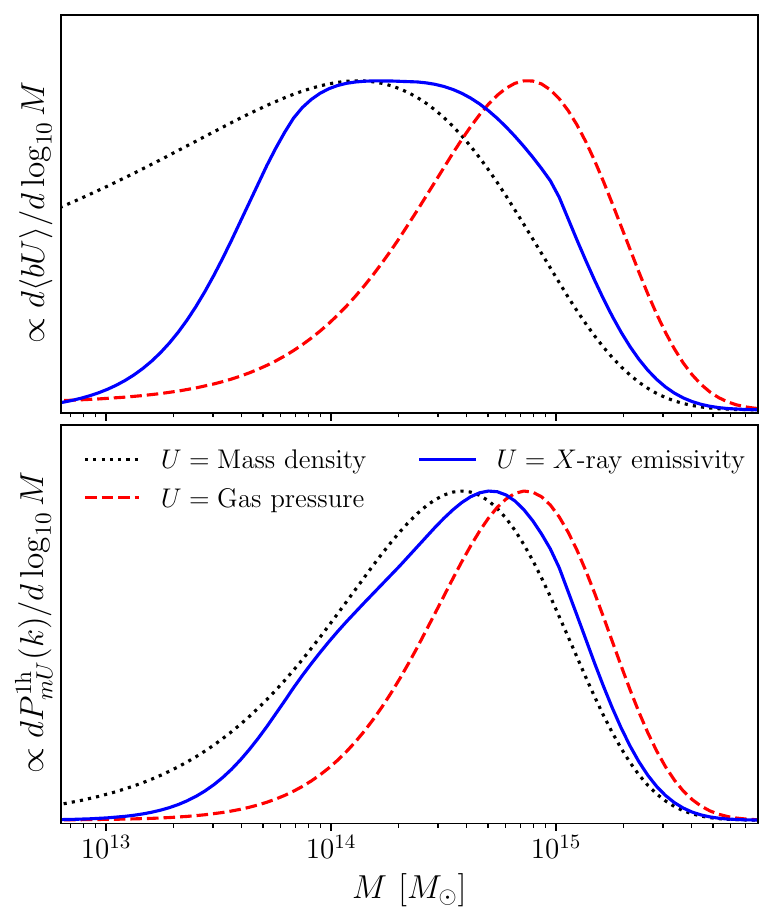}
        \caption{{\sl Top:} relative contribution from different halo masses to the bias weighted average $\langle bU\rangle$, which dominates the amplitude of the power spectrum on large scales. {\sl Bottom:} relative contribution from different halo masses to the 1-halo power spectrum, which dominates the amplitude on small scales, at $k=1\,{\rm Mpc}^{-1}$. Results are shown at $z=0$ for the matter power spectrum (dotted black), for the matter-pressure cross correlation (dashed red), and for the correlation between matter and \xray emissivity (solid blue). On small-scales, where baryonic effects are most relevant, cross-correlations between cosmic shear and \xray emission from hot gas traces almost the same mass scales that the matter power spectrum is sensitive to.}\label{fig:mass_contributions}
    \end{figure}
    Weak lensing distorts the shapes of background galaxies, inducing correlations between them and with the intervening large-scale structure. The resulting ``cosmic shear'' $\gamma$ is a spin-2 field with vanishing $B$-modes at leading order \citep{astro-ph/9912508} (see \cite{1967JMP.....8.2155G} for further details about spin quantities defined on the sphere).
    
    The $E$-mode component of the shear field is a line-of-sight integral with the form of Eq. \ref{eq:projfield}, with the three-dimensional field corresponding to the matter overdensity field\footnote{Strictly speaking, the relation between the shear $E$-mode and the matter overdensity involves an additional scale-dependent factor in harmonic space of the form $f_\ell=\sqrt{(\ell+2)(\ell+1)\ell(\ell-1)}/(\ell+1/2)^2$, which is nevertheless negligibly close to 1 on the scales explored here.} $\delta_m({\bf x},z)$, and the radial kernel
    \begin{equation}
      q_\gamma(\chi)=\frac{3}{2}\Omega_m\,H_0^2(1+z)\chi\int_z^\infty\,dz'\,p(z')\,\frac{\chi(z')-\chi}{\chi(z')}.
    \end{equation}
    Here, $H_0$ is the expansion rate today, and $p(z)$ is the redshift distribution of source galaxies. \al{Following~\cite{2309.11129}, we neglected the impact of intrinsic alignments (IAs) on the cross-correlations studied in this article. Ideally, the contribution from IAs would be constrained self-consistently by jointly analysing the $\gamma X$ and $\gamma y$ correlations in combination with the cosmic shear power spectra. We leave this study for future work. It is worth noting that the DES Y3 analysis~\cite{2105.13544} did not find strong evidence for IAs. We will also neglect the impact of photometric redshift uncertainties, as well as multiplicative shape measurement bias. Given the relatively tight calibration priors on these systematics, their impact on the astrophysical constraints presented here should be small.}

    To understand the value of the different cross-correlations studied in this work as probes of the cosmic baryon component in the context of cosmological weak lensing analysis, it is useful to quantify the halo mass scales that they are sensitive to. Figure \ref{fig:mass_contributions} shows the contribution to different power spectra from different halo masses. The contribution to the bias-weighted average of a quantity $U$ is shown in the top panel (i.e. the integrand in Eq. \ref{eq:bU}), while the bottom panel shows the contribution to the cross-spectrum between the matter overdensity and $U$ in the 1-halo regime (i.e. the integrand in Eq. \ref{eq:pk1h}). Results are shown for the matter power spectrum (dotted black), the mass-pressure power spectrum (dashed red), and the correlation between matter and \xray emissivity (solid blue). The volume integral of the thermal gas pressure scales as $\sim M^{5/3}$ approximately, due to the dependence on gas temperature, and the tSZ cross-correlation (represented here by the matter-pressure power spectrum) is thus sensitive to larger masses than the matter power spectrum. In turn, the \xray emissivity has a much milder dependence on temperature (as long as the gas is hot enough to produce significant bremsstrahlung emission), and thus scales approximately as $\sim M$. Cross-correlations with the \xray emission from hot gas can therefore be used effectively to study the impact of baryonic physics in the matter power spectrum, since they track a similar range of halo masses. \al{We did not consider the effect of baryons in the modelling of cosmic shear in our fiducial analysis. We tested that, including the impact of baryonic effects by modelling the gas content of dark-matter haloes as described in Section \ref{ssec:meth.halomodel} led to only small changes to our final constraints, shifting the central parameter values by $\sim 0.5\sigma$ on average.}

\subsection{Power spectra and covariances}\label{ssec:meth.cls}
  We estimate all power spectra using the pseudo-$C_\ell$ algorithm \citep{astro-ph/0105302} as implemented in \nmt\footnote{\url{https://github.com/LSSTDESC/NaMaster}} \citep{1809.09603}. The method is based on the analytical calculation of the statistical coupling between different power spectrum multipoles (summarised in the so-called mode-coupling matrix) caused by the presence of a sky mask. The details of the method, including its application to spin-2 fields (as is the case for cosmic shear) are described in \cite{1809.09603}.

  A key peculiarity of cosmic shear data is the fact that the field is only defined at the positions of galaxies in the catalog. Effectively this implies that a close-to-optimal sky mask in this case is given by the sum of shape weights for all sources in each pixel\footnote{Note that substantial progress has been made recently in the analysis of discretely-sampled fields \citep{2312.12285,2407.21013}. Although we use a pixel-based approach, this has been extensively validated for the shear sample studied here (see e.g. \citep{2203.07128,2403.13794}). See also \cite{2411.15063}.}. Such a complex mask leads to enhanced residual mode-coupling that must be propagated to the theoretical predictions, convolving the predicted $C_\ell$s with the associated bandpower window functions. For further details regarding the treatment of cosmic shear power spectra, see \cite{2010.09717}.

  The covariance matrix of the estimated power spectra was calculated analytically. Specifically, we calculate the disconnected part of the ``Gaussian'' covariance using the improved Narrow Kernel Approximation, described in \cite{1906.11765,2010.09717}, incorporating the impact of mode-coupling via approximate pseudo-$C_\ell$-like methods. We include the non-Gaussian contribution to the covariance matrix due to the 1-halo trispectrum \citep{astro-ph/0205468}, which is mostly relevant for the $Xy$ cross-correlation (which itself is dominated by the 1-halo term of low-redshift clusters). Other non-Gaussian contributions, particularly super-sample covariance terms, were found to be small for shear-$y$ cross-correlations in \cite{2109.04458}, and for shear-\xray correlations in \cite{2309.11129}.
  
  All maps used to estimate power spectra were constructed using the \hpx\footnote{\url{http://healpix.sourceforge.net/}}\cite{astro-ph/0409513} pixelisation scheme with a resolution parameter $N_{\rm side}=1024$, corresponding to $\sim3.4'$ pixels. The choices made to construct these maps and their masks are described for each individual dataset in Section \ref{sec:data}. Power spectra were measured in $\ell$ bins with linear spacing of $\delta\ell=30$ for $\ell<240$, and logarithmic bins with $\delta\log_{10}\ell=0.055$ in the range $240<\ell<3N_{\rm side}$. In our analysis, we use scales with bin centers in the range $30<\ell<2000$ (the exact scale cuts are discussed in Section \ref{sec:res}), resulting in 24 bandpowers per $C_\ell$.

\section{Data}\label{sec:data}

\subsection{DES}\label{ssec:data.des}
  We use cosmic shear data from the 3-year data release of the Dark Energy Survey \citep{2101.05765} (DES-Y3). The DES \citep{astro-ph/0510346} is an imaging galaxy survey covering $\sim5000\,{\rm deg}^2$ of the southern sky. Observations were taken from the Cerro Tololo International Observatory with the 4-meter Blanco telescope. We use the publicly available 3-year cosmic shear sample \citep{2011.03408}, containing more than $10^8$ sources over an effective area of $4143\,{\rm deg}^2$, for a number density $\bar{n}_{\rm eff}=5.6\,{\rm arcmin}^{-2}$.

  These data were analysed, as described in \cite{2403.13794} and \cite{2309.11129}, following closely the steps described in the official DES analysis \citep{2011.03408,2203.07128}. The sample was divided into four redshift bins, spanning the range of photometric redshifts $z_{\rm ph}<1.5$. The official redshift distributions were used to interpret the signal from each of these bins. The mean ellipticity in each bin was subtracted from all sources, and the resulting ellipticities were corrected for the mean multiplicative bias estimated from the {\sc Metacalibration} response tensor. We then construct spin-2 shear maps by binning the galaxy ellipticities onto pixels, using the \hpx pixelisation scheme with resolution $N_{\rm side}=1024$ (corresponding to a pixel resolution of $\delta\theta\simeq3.4'$).

\subsection{ROSAT}\label{ssec:data.x}
  We use \xray data from the final data release of the ROSAT All-Sky Survey (RASS) \citep{astro-ph/9909315}. ROSAT observed the full \xray sky in the soft band ($0.1\,{\rm keV}\lesssim\energ\lesssim2.0\,{\rm keV}$), with modest energy resolution ($\delta \energ/\energ\sim0.4$). We make use of data collected during the six-month all-sky survey phase of the mission, until the main detector (PSPC-C) was destroyed due to a pointing glitch. We use the photon catalog as provided by the German Astrophysics Virtual Observatory \cite{vo:rosat_photons}, selecting only photons in the observed energy range $\energm\in[0.5,2]\,{\rm keV}$. We also use the exposure images available with the RASS release, reprojecting them onto \hpx pixelisation to generate a full-sky exposure map with the same resolution used for the cosmic shear maps ($N_{\rm side}=1024$). We then construct a photon count rate map by binning all photons in the desired energy range onto these pixels, and dividing by the exposure map and associated pixel area. When interpreting the ROSAT data, we account for the effective point-spread function of the instrument, modeling it as a Gaussian with a full-width half-max (FWHM) spread of 1.8 arcminutes \citep{1997ApJ...485..125S}.

  We also use data from the Second RASS Point Source Catalog (2RXS) \citep{1609.09244} to quantify the impact of point sources (predominantly \xray AGN) in our measurements. We select sources with fluxes above $0.02$ photons per second, masking all pixels containing such sources\footnote{More in detail, we produce a point-source mask at high resolution ($N_{\rm side}=2048$), roughly matching the instrument PSF, and then downgrade the resulting binary map to our base resolution.}. In addition to this, we mask out all regions of the sky with exposures below 100s, and applied a Galactic mask that removes areas of large Milky Way dust emission (see details in \cite{koukoufilippas2019}), to reduce potential contamination from Galactic foregrounds.

\subsection{Compton-$y$ maps}\label{ssec:data.y}
  We use maps of the Compton-$y$ parameter constructed from the multi-frequency \planck maps. Our baseline $y$ map is the \planck 2015 full-sky map, generated using the MILCA component separation technique~\cite{1502.01596}. MILCA is a modified version of the internal linear combination (ILC) technique, which extracts signals with a known spectrum from optimally-designed linear combinations of multi-frequency maps. The linear weights are often dependent on angular scale and sky position. Details regarding the specific implementation used can be found in \cite{1502.01596} and \cite{1007.1149}.

  We compare the results found with this fiducial map against two other public $y$ maps. First, we use the map constructed from the final \planck data release ~\cite{planck_npipe} (PR4) using a needlet ILC (NILC) approach. The map exhibits lower noise levels and a reduced level of contamination from Galactic and extragalactic dust emission. To further test for the potential impact of contamination from Galactic and extragalactic foregrounds, we also employ the $y$ maps released by \cite{mccarthy2024} from the PR4 dataset. In this case, various versions of the $y$ map were generated using a constrained ILC method that deprojects contamination from other sky components with known spectra. In particular, we employ the $y$ maps with no contaminant deprojection, as well as those with CMB and CMB $+$ CIB deprojected.
  
  To minimise contamination from Galactic foreground and extragalactic point sources we construct a mask using the \planck Galactic mask covering $60\%$ of the sky, as well as the point-source mask from~\cite{1303.5088,1502.01596}.

\section{Results}\label{sec:res}

\subsection{Power spectrum measurements}\label{ssec:res.cls}
  \begin{figure*}[htpb]
    \centering
    \includegraphics[width=\textwidth]{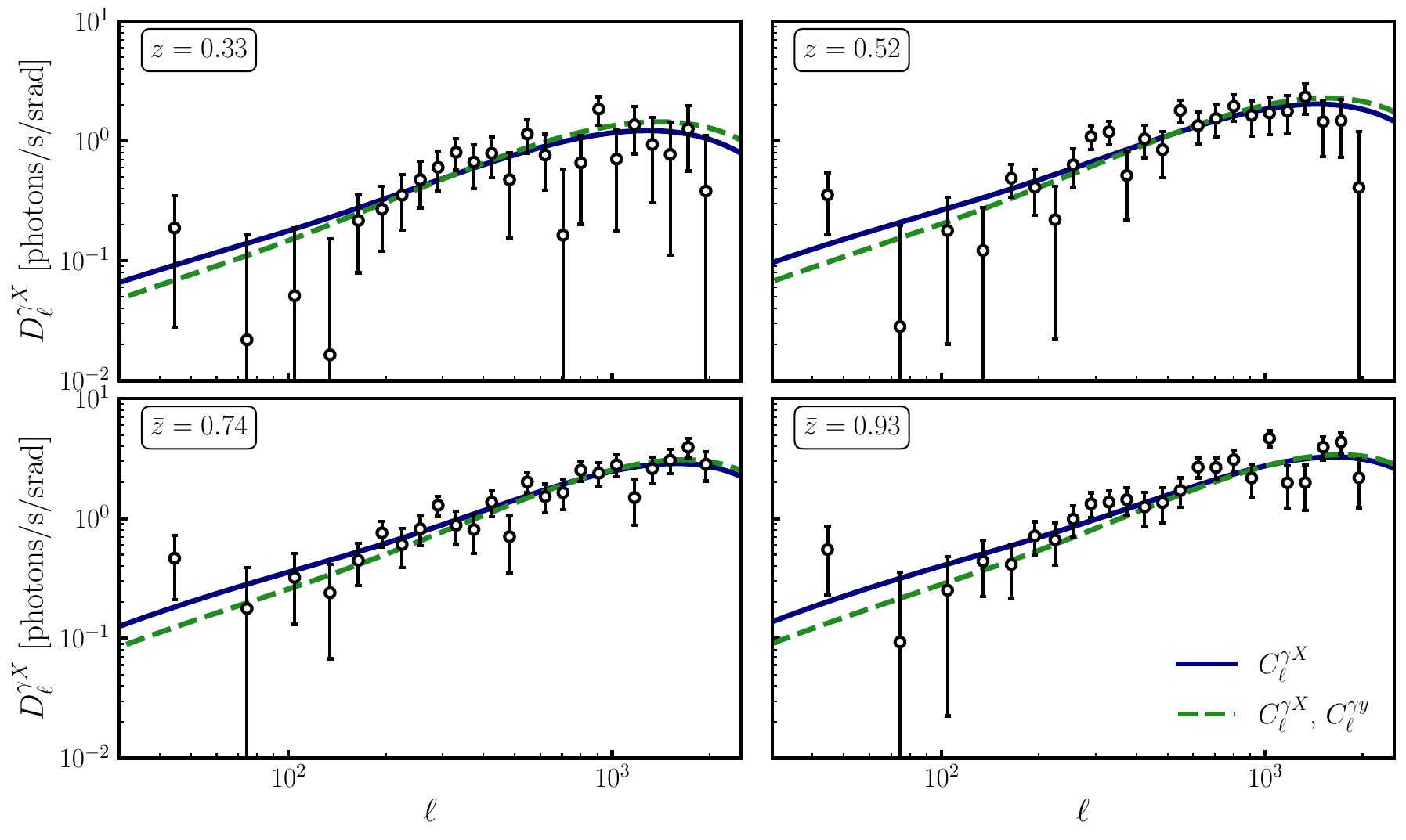}
    \includegraphics[width=\textwidth]{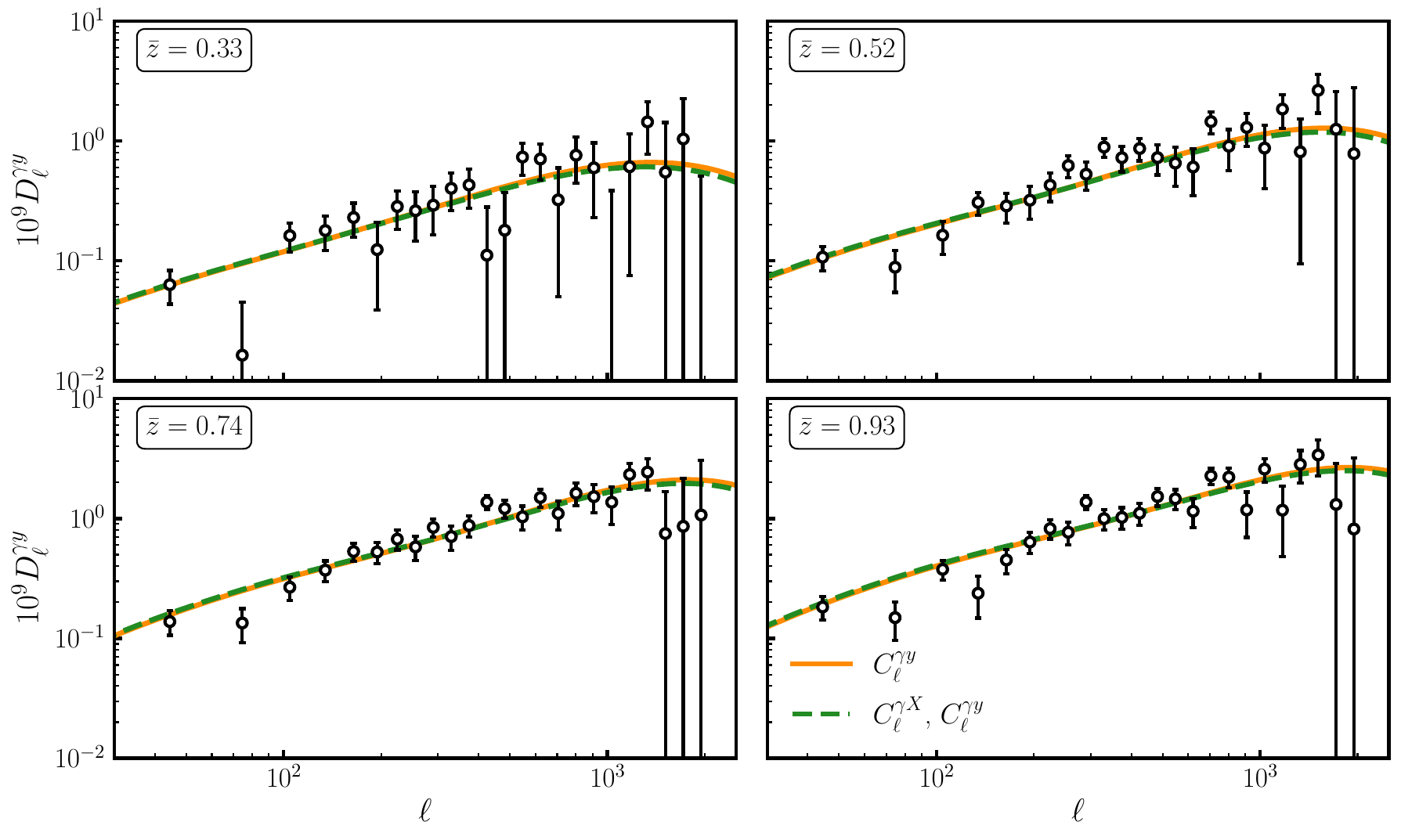}
    \caption{Angular cross-power spectra of the 4 DESY3 cosmic shear redshift bins with the ROSAT \xray map (top panels) and the \planck 2015 Compton-$y$ map (bottom panels). We show theory predictions from the bestfit to $C_\ell^{\gamma X}$ (blue), $C_\ell^{\gamma y}$ (orange) and their combination (dashed green) where individual $\chi^2$ are listed in Table~\ref{tab:snr_det}. For visualisation purpose, we used $D_\ell=\ell(\ell+1)C_\ell/2\pi$}.
    \label{fig:shear_x_y_ps}
  \end{figure*}
  The top and bottom panels of Figure  \ref{fig:shear_x_y_ps} show the angular cross-correlations of DES cosmic shear data with the ROSAT \xray maps and the \planck Compton-$y$ map cosmic shear, respectively. Results are shown for the four different DES redshift bins. The different theoretical predictions shown will be described in the next sections. The data analysis pipeline used to construct these measurements, including map construction for all probes, as well as power spectrum and covariance matrix estimation, was developed as part of the {\tt Cosmotheka}\footnote{\url{https://github.com/Cosmotheka/Cosmotheka}} package~\cite{2105.12108}. All cross-correlations are detected at high significance, with signal-to-noise ($S/N$) ratios above 10 for each individual cross-correlation. Both cross-spectra ($\gamma\times X$ and $\gamma\times y$) are detected with comparable significance (see Table \ref{tab:snr_det}), with the Compton-$y$ cross-correlation achieving consistently higher $S/N$.

  \begin{table*}[t!]
    \begin{ruledtabular}
      \begin{tabular}{lccc|ccc}
        &\multicolumn{3}{c|}{\textbf{DES} $\times$ \textbf{ROSAT}} &\multicolumn{3}{c}{\textbf{DES} $\times$ \textbf{Planck tSZ}} \\
        {\bf Bin} & \textbf{$E$-mode SNR} & \textbf{$B$-mode} ${\bm \chi^2}$ & \textbf{Best-fit}
        ${\bm \chi^2}$ \textbf{(PTE)} & \textbf{$E$-mode SNR} & \textbf{$B$-mode} ${\bm \chi^2}$ & \textbf{Best-fit} ${\bm \chi^2}$ \textbf{(PTE)} \\
        \hline
        $\bf z_1$ & 10.0 & 14.6 (0.93) & 19.9 (0.71) & 11.1 & 22.4 (0.56) & 23.5 (0.49) \\
        $\bf z_2$ & 14.2 & 17.6 (0.82) & 25.8 (0.36) & 17.5 & 28.5 (0.24) & 29.1 (0.22) \\
        $\bf z_3$ & 18.3 & 23.1 (0.51) & 22.8 (0.53) & 22.4 & 16.0 (0.89) & 27.6 (0.28) \\
        $\bf z_4$ & 18.0 & 24.4 (0.44) & 30.4 (0.17) & 22.4 & 20.0 (0.70) & 46.4 (0.004) \\
      \end{tabular}
    \end{ruledtabular}
    \caption{Detection significance of the correlation between ROSAT \xray maps or \planck tSZ maps with DES cosmic shear E-modes expressed as a signal-to-noise ratio (SNR). The table also displays the consistency of the correlation measured from cosmic shear B-modes with zero via a $\chi^2$ and the associated probability to exceed (PTE). We also show the $\chi^2$ for the best fit model to the $\gamma X$ + $\gamma Y$ data vector. Note that, although the cross-correlation of the highest redshift bin with the tSZ map achieves a relatively large $\chi^2$, the distribution of $\chi^2$ values for all cross-correlations listed in the table is compatible with a $\chi^2$ distribution with the same number of degrees of freedom (see details in the main text).}\label{tab:snr_det}
  \end{table*}

  In addition to the cross-correlation with the shear $E$-modes, we also measure the $B$-mode power spectra. The $\chi^2$ and associated probability-to-exceed (PTE) values of all spectra are reported in Table \ref{tab:snr_det}. \al{The PTE values reported in the table were computed assuming that the calculated $\chi^2$ follows a chi-squared distribution with a number of degrees of freedom $N_{\rm dof}$ given by the number of data points. This is appropriate, as the best-fit model used here is that found by combining all the $C_\ell^{\gamma X}$ and $C_\ell^{\gamma y}$ measurements, and not just the individual power spectra listed in the table. Nevertheless, reducing this $N_{\rm dof}$ by up to 3 (the total number of free parameter in our baseline model) would not change the qualitative conclusions drawn here.} As expected, the measurements are consistent with zero, with all PTEs in the range $>0.24$. This confirms that no systematics are present in the cosmic shear data that could lead to a $B$-mode signal correlated with the large-scale structure. It also serves as an indirect confirmation that our covariance matrix is reasonable, since a significant under- or over-estimate of the covariance would have led to PTE values that are suspiciously close to 0 or 1, respectively. Finally, we note that, as reported in \cite{2403.13794}, the same analysis pipeline is able to recover the official DES cosmic shear auto-correlations (and their covariance), presented in \cite{2203.07128}, at high accuracy. We make these power spectrum measurements publicly available\footnote{\url{https://github.com/Cosmotheka/Cosmotheka_likelihoods/tree/main/papers/syx}}.
  
\subsection{Constraints on gas thermodynamics from $\gamma X$ + $\gamma y$}\label{ssec:res.params}
  We model the gas distribution in halos following Section~\ref{ssec:meth.halomodel}, using the halo model to compute the predicted angular power spectrum. To extract constraints on the parameters of the model we use a Gaussian likelihood for the data such that
  \begin{equation}
    \log p(\hat{\bf C}|\vec{\theta}) = - \frac{1}{2} \left[ \hat{\bf C} - {\bf C}_\mathrm{th}(\vec{\theta}) \right]^\top \mathbf{\Sigma}^{-1} \left[ \hat{\bf C} - {\bf C}_\mathrm{th}(\vec{\theta}) \right],
  \end{equation}
  where $\hat{\bf C}$ is our data vector of measured power spectra, ${\bf C}_{\rm th}(\vec{\theta})$ is the theoretical prediction, dependent on parameters $\vec{\theta}$, and $\mathbf{\Sigma}$ is the covariance matrix of these measurements. We will consider data vectors constituted by different combinations of the cross-correlations between cosmic shear, \xray, and $y$ maps $\{C_\ell^{\gamma X},\,C_\ell^{\gamma y},\,C_\ell^{Xy}\}$. Unless stated otherwise, for every dataset we use multipoles from $\ell=30$ to $\ell=2073$, corresponding to $24$ multipole bins. We discuss the stability of our results with respect to tSZ scale cuts in Section~\ref{sssec:res.params.tszsys}.

  In the rest of this section we will present constraints on the free parameters of the model. In the simplest scenario these will include $\{\lMc, \Gamma, \alpha_T\}$, but we will discuss the ability of this model to simultaneously describe $C_\ell^{\gamma X}$ and $C_\ell^{\gamma y}$, and the results obtained when extending the model to include the impact of uncertainties in non-thermal gas pressure and \xray AGN contamination. In our analysis we fix all cosmological parameters to the best-fit values found by \planck \cite{1807.06209}.

  \subsubsection{Constraints on a minimal hydrodynamic model}\label{sssec:res.params.minimal}
\begin{figure}[t!]
    \centering
    \includegraphics[width=\columnwidth]{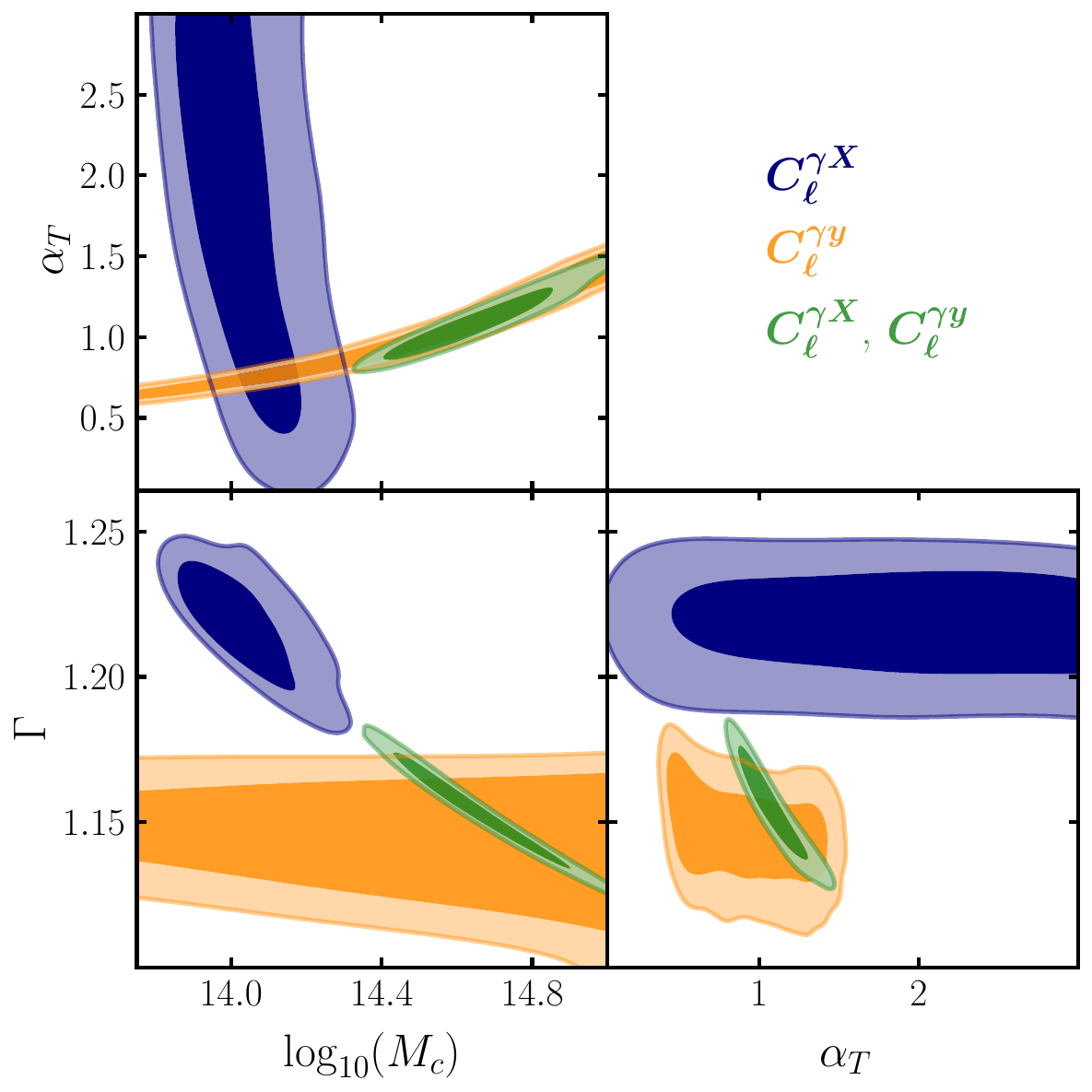}
    \caption{2D marginalised posterior distributions for our minimal hydrodynamic model, derived from the \xray-shear correlation (blue), from the tSZ-shear correlation (orange), and from their combination (green). Angular power spectrum predictions from best-fit models of these three data combinations are shown in Fig.~\ref{fig:shear_x_y_ps}. Note that the constraint on $\alpha_T$ from the shear-tSZ correlation is fully driven by the prior bounds on $\mathrm{log}_{10}(M_c)$. \al{Combined constraints are pulled towards lower $\Gamma$ values due to the higher SNR of the tSZ-shear correlation, following the \xray-shear degeneracy line.}}
    \label{fig:baseline_contours}
    \end{figure}

    We start by constraining a minimal hydrodynamic halo model using $C_\ell^{\gamma X}$, $C_\ell^{\gamma y}$, and the combination of both. In this case, the model is that described in Section \ref{ssec:meth.halomodel}, with only three free parameters: the halfway mass $\lMc$, the polytropic index $\Gamma$, and the deviation from virial temperature parametrised by $\alpha_T$. As done in past analyses (e.g. \cite{1712.05463}), we do not attempt to correct for the impact of AGN contamination beyond masking the resolved point sources from the 2RXS catalog. The two-dimensional marginalised posterior distributions for this case are shown in Fig.~\ref{fig:baseline_contours}.

    Considering first the constraints found from $\gamma\times X$ alone (blue contours), we see that, as found in \cite{2309.11129}, the \xray cross-correlation alone is in principle able to jointly constrain $\lMc$ and $\Gamma$, which control the density and scale dependence of the bound gas, but is only able to place a weak lower bound on the temperature parameter $\alpha_T$. This is not unexpected, given the relatively mild dependence of the \xray intensity on the plasma temperature. In contrast, the strong degeneracy between temperature and density makes it impossible for the tSZ cross-correlation alone (orange contours) to constrain $\lMc$ and $\alpha_T$ simultaneously, although the measurement is highly sensitive to a particular combination of them. The degeneracy line is well approximated by
    \begin{equation}
      (\alpha_T-\alpha_0)\left(\frac{M_c}{M_{c,0}}\right)^\lambda\simeq 1,
    \end{equation}
    with $\log_{10}M_{c,0}=14.97$, $\alpha_0=0.43$, and $\lambda=-0.55$, and we find that the tSZ data alone can place constraints on this particular combination at the level of a few percent:
    \begin{equation}\label{eq:alphat}
      \tilde{\alpha}_T\equiv\alpha_T\left[\alpha_0 + \left(\frac{M_c}{M_{c,0}}\right)^{-\lambda}\right]^{-1} = 1.003\pm 0.043.
    \end{equation}
    At the same time, the $\gamma y$ correlation is directly sensitive to $\Gamma$ through the scale dependence of the gas profile. Quantitatively, we find the following constraints on $\Gamma$ from each individual probe:
    \begin{align*}
      \Gamma &= 1.218^{+0.016}_{-0.011} \quad (C_\ell^{\gamma X}), \\
      \Gamma &= 1.147^{+0.015}_{-0.013} \quad (C_\ell^{\gamma y}).
    \end{align*}
    These constraints are in tension at the $\sim3.5\sigma$ level. Furthermore, estimating the suspiciousness statistic~\cite{1910.07820, 2007.08496} of both datasets, we find them to be in tension at the $2.8\sigma$ level.

    This level of tension would seemingly make it impossible to interpret both sets of cross correlations simultaneously within this model, even though doing so allows us to break their individual parameter degeneracies. Nevertheless, ignoring this tension for the moment (we will return to it shortly), a joint analysis of both $C_\ell^{\gamma X}$ and $C_\ell^{\gamma y}$ leads to the following parameter constraints (shown as green contours in Fig. \ref{fig:baseline_contours}):
    \begin{align}\nonumber
      \lMc &= 14.66^{+0.12}_{-0.19},\\\label{eq:param_minim}
      \Gamma &= 1.154^{+0.015}_{-0.012},\\\nonumber
      \alpha_T &= 1.10^{+0.11}_{-0.19}.
    \end{align}
    Interestingly, in spite of the tension between both datasets within this model, the joint best fit parameters are able to describe the joint dataset reasonably well. The best-fit chi-squared statistic for the full data vector is $\chi^2=204.0/189$, corresponding to a PTE of $22\%$\footnote{\al{Note that this PTE was estimated for a number of degrees of freedom given by $N_{\rm dof}=N_{\rm data}-N_{\rm param}$, where $N_{\rm param}=3$ is the number of free parameters in our baseline model. For models with non-linear parameter depedencies, as is the case here, some care should be exercised when estimating the effective number of parameters constrained by the data (see e.g. \cite{2007.01844}). However, given the small number of parameters contained by our model, our conclusions regarding the ability of the best-fit model to describe the data is not affected by this.}}. The corresponding $\chi^2$ values for the individual power spectra, and their PTEs, are listed in Table \ref{tab:snr_det}. We obtain acceptable PTEs for all spectra except the cross-correlation between the third cosmic shear bin and the tSZ map. This PTE improves slightly when using one of the alternative tSZ maps discussed in Section \ref{sssec:res.params.tszsys} (MC23 -- ${\rm PTE}=0.06$). However, it is not necessarily surprising to have a small number of anomalously low PTEs out of a sample of several data vectors. Quantitatively, a Kolmogorov-Smirnoff test comparing the $\chi^2$ values listed in this table against a $\chi^2$ distribution with 24 degrees of freedom yields a $p$-value of 0.24 (or 0.9 when including both $E$- and $B$-mode power spectra).

    The best-fit model predictions are shown as green lines in Figure \ref{fig:shear_x_y_ps}. The figure also shows the best-fit predictions found for each set of cross-correlations when analysing them individually. Note that, in all cases, we obtain reasonable best-fit PTEs, neither suspiciously low, nor high. Hence the good $\chi^2$ achieved by the joint best fit model, in spite of the parameter tension between both datasets, is not driven by an overestimate of the statistical uncertainties.

    The next sections explore three different ways to eliminate this tension and reconcile both datasets within the same model.
    \begin{figure}[t!]
      \centering
      \includegraphics[width=\columnwidth]{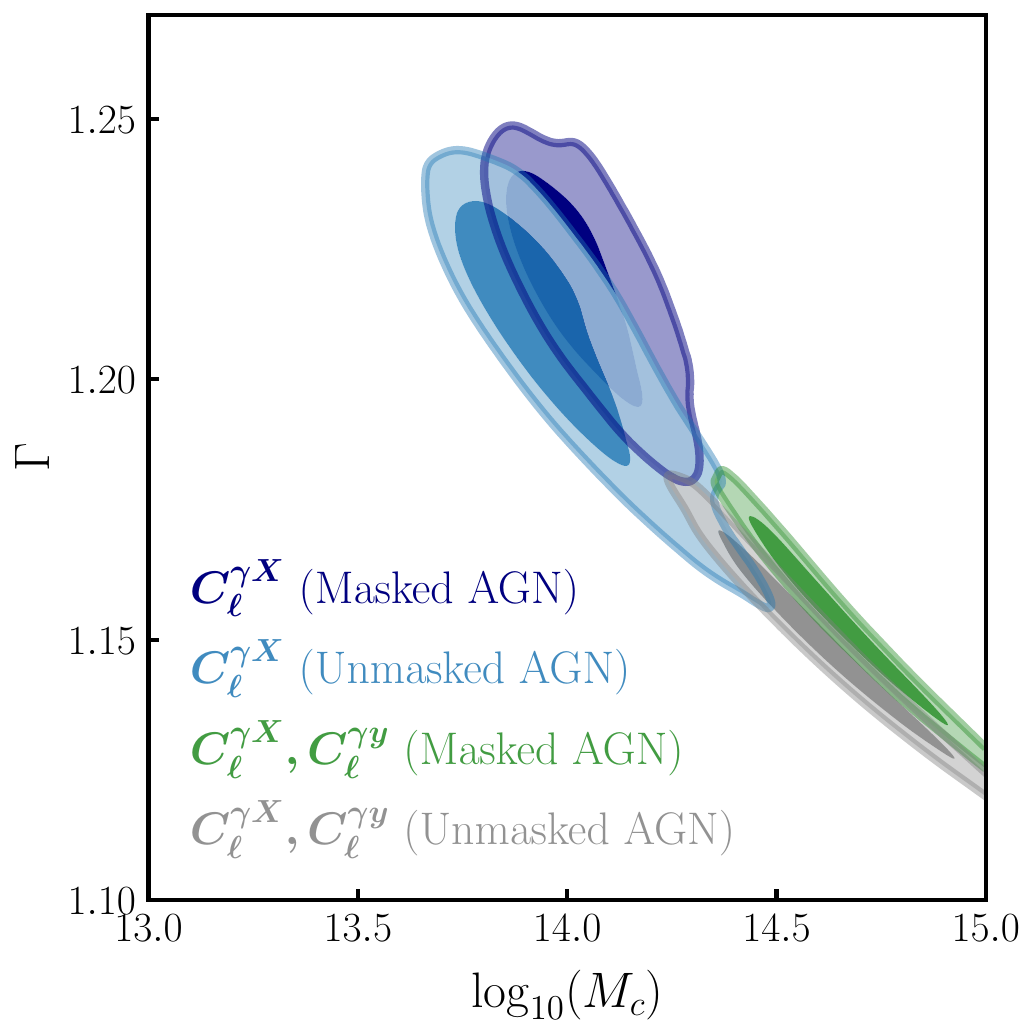}
      \caption{2D marginalised posterior distributions for $C_\ell^{\gamma X}$ alone and in combination with $C_\ell^{\gamma y}$ power spectra masking the brightest AGN (darkblue, green) or without masking it (lightblue, gray). The impact of our choice of masking brightest AGN is not significant enough to explain the discrepancy on the inferred polytropic index $\Gamma$.}
      \label{fig:mask_agn}
    \end{figure}

  \subsubsection{AGN contamination}\label{sssec:res.params.agn}

    Emission from unresolved non-thermal sources, mostly AGNs, acts as a significant contaminant to the \xray emission from diffuse IGM gas. In this section, we examine the impact of this contamination on our constraints, explore ways to account for it or mitigate it, and show that, doing so, we are able to describe our two datasets within a single model without tension.

    The baseline results presented in the previous subsection used \xray maps from ROSAT after masking out the brightest \xray point sources from the RASS 2RXS source catalog~\cite{1609.09244}, using a flux cut of $0.02$ photons$/\mathrm{s}$. This masking leads to an overall reduction in the amplitude of all power spectra involving ROSAT (both in cross-correlation with cosmic shear and tSZ, the latter of which we will discuss in Section \ref{ssec:res.extern}) by between 10 and 20\%. Note that this is not necessarily representative of the pure impact of contamination from the correlation of these sources with the large-scale structure: many of these AGNs reside at the centre of massive haloes that also emit a significant amount of bremsstrahlung radiation. Thus, masking these sources also removes part of the signal we are trying to measure. Both of these effects (the removal of genuine contamination from AGNs and the undesired subtraction of part of the gas signal) lead to a reduction in the amplitude of the \xray cross-correlations.

    To quantify the impact of these effects, we repeat our analysis using power spectra measured with no \xray point source masking. The result of this exercise is shown in Fig. \ref{fig:mask_agn}. When considering $C_\ell^{\gamma X}$ alone, removing the point source mask leads to a reduction in the inferred value of both $\lMc$ and $\Gamma$. Lowering $\lMc$ leads to higher gas densities in the less massive halos, while lowering $\Gamma$ leads to steeper profiles (and hence a larger density), and both therefore enhance the \xray signal. Nevertheless, as shown in the figure, this shift is not sufficient to resolve the apparent tension between both datasets.

    In addition to the detected point sources we mask, the majority of \xray-emitting AGNs are unresolved \citep{astro-ph/0501058}, and may constitute a significant contaminant to \xray cross-correlations. To quantify the impact of this unresolved component, we construct a theoretical estimate of their contribution. We do so making use of the \xray luminosity function model \cite{1503.01120}, and following the procedure outlined in Section \ref{sssec:meth.xys.x} and Appendix \ref{app:agn}. Since the model used to generate this prediction has large uncertainties (e.g. in the small-scale clustering of galaxies, or the role of AGN obscuration in the soft band), we also attempt to marginalise over them. Concretely, we model any cross-correlation of the \xray map with another tracer $u$ as
    \begin{equation}
      C_\ell^{uX}=C_\ell^{uX,{\rm gas}}+A_{\rm AGN}C_\ell^{uX,{\rm AGN}},
    \end{equation}
    where $C_\ell^{uX,{\rm gas}}$ and $C_\ell^{uX,{\rm AGN}}$ are the models for the gas and unresolved AGN contributions, respectively (with the former depending on the hydrodynamical parameters). We predict the unresolved contribution to the cross-correlation with shear and tSZ ($C_\ell^{\gamma X,{\rm AGN}}$ and $C_\ell^{yX,{\rm AGN}}$) to be around 20\%, as shown in Fig. \ref{fig:cls_agn}. The amplitude $A_{\rm AGN}$ is a free parameter that we marginalise over, assuming a Gaussian prior with mean $\bar{A}_{\rm AGN}=1$ (corresponding to our predicted AGN contribution) and standard deviation $\sigma(A_{\rm AGN})=1$. Effectively we thus allow the unresolved AGN contribution to be 100\% larger or smaller than our expectation, within 1-$\sigma$.

    \begin{figure}
        \centering
        \includegraphics[width=0.9\columnwidth]{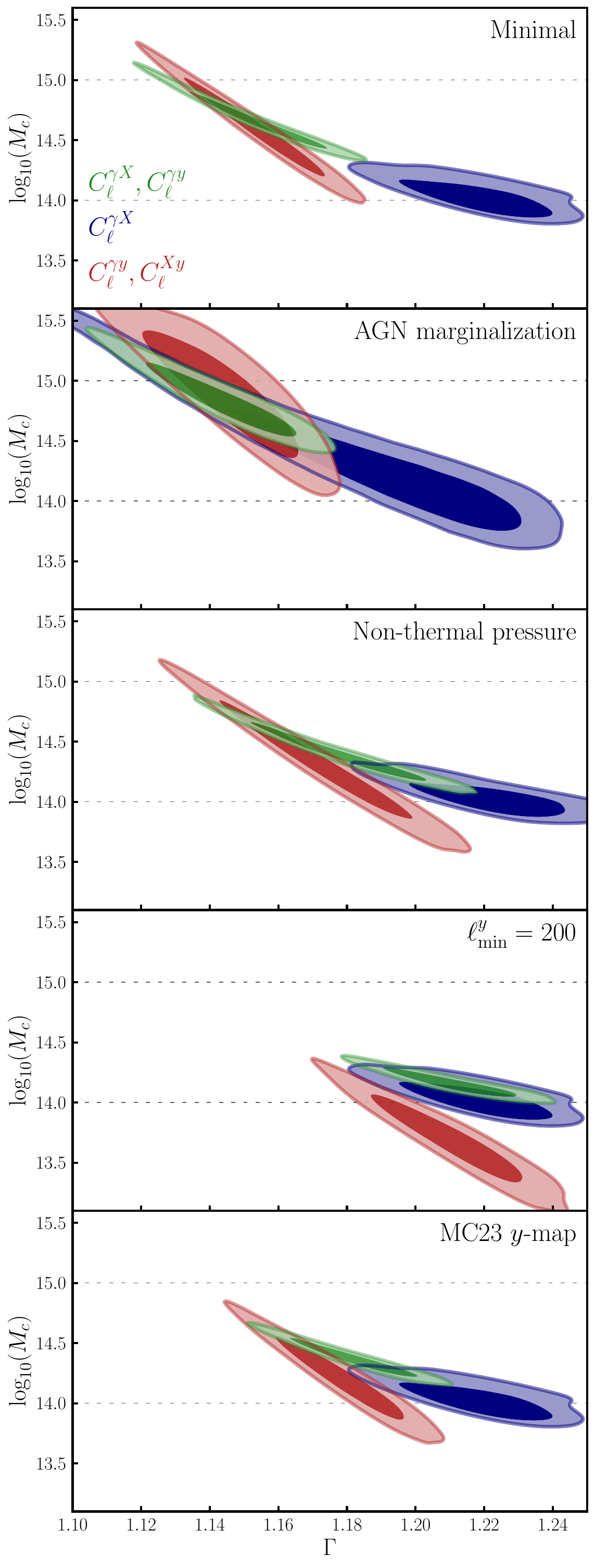}
        \caption{2D marginalised posterior distributions in the $\Gamma - \mathrm{log}_{10}(M_c)$ plane for different analysis settings and data combinations. This figure highlights our most successful attempt at reconciling parameter constraints from the cross correlations studied in this article: marginalising over AGN contamination in \xray data.}
        \label{fig:gamma-MC-constraints}
    \end{figure} 

    The resulting constraints on \al{$\lMc$ and $\Gamma$} are shown in the \al{second} panel of Figure~\ref{fig:gamma-MC-constraints} for $C_\ell^{\gamma X}$ (blue) and $C_\ell^{\gamma X}+C_\ell^{\gamma y}$ (green). Accounting for the unresolved contribution, and marginalising over its uncertainties allows models with larger $\lMc$ and, importantly, smaller $\Gamma$ to describe the $\gamma$-$X$ correlation. This significantly increases the overlap in parameter space between the individual $\gamma X$ and $\gamma y$ constraints, eliminating the tension between them altogether. The additional freedom in the model leads to an increase in posterior parameter uncertainties. Combining both datasets, we obtain the following constraints on the hydrodynamic parameters:
    \begin{align*}
      \lMc &= 14.83^{+0.16}_{-0.23},\\
      \Gamma &= 1.144^{+0.016}_{-0.013},\\
      \alpha_T &= 1.30^{+0.15}_{-0.28}.
    \end{align*}
    These correspond to a $\sim1\sigma$ increase in $\lMc$, and a less significant increase in $\alpha_T$ from the constraints found with our minimal model (Eq. \ref{eq:param_minim}). The value of $\Gamma$ obtained is in line with the value preferred by the tSZ cross-correlation.

    \al{Additionally, given the good agreement in parameter space of $\gamma X$, $\gamma y$ and $Xy$ constraints, one could further improve the constraints on gas distribution and thermodynamics including all spectra in the data vector. We report on the marginalised posterior contours in Appendix~\ref{app:full_combo} and find 
    \begin{align*}
        \lMc &= 14.69^{+0.09}_{-0.10},\\
        \Gamma &= 1.155 \pm 0.008,\\
        \alpha_T &= 1.13^{+0.08}_{-0.10},
    \end{align*}
    with a major improvement on the $\alpha_T$ 68\% errors, still consistent with the virial relation.}

  \subsubsection{Non-thermal pressure}\label{sssec:res.params.nonth}

    Our minimal hydrodynamic model assumes the total gas pressure to be dominated by thermal pressure, which allows us to univocally relate the halo density and temperature profiles (and, from them, predict the tSZ and \xray signals). Since the tSZ signal is significantly more sensitive to the gas temperature, allowing for non-thermal pressure will modify the region of parameter space favoured by the $\gamma y$ cross-correlations, potentially bringing it into better agreement with the constraints derived from $C_\ell^{\gamma X}$. To explore this possibility, we extend the minimal model by freeing the non-thermal pressure parameter $\gamma_T$ (see Eq.~\ref{eq:nt_pres_definition}), which modifies the shape of the bound gas temperature profile. We marginalise over this parameter imposing a Gaussian prior $\mathcal{N}(1.5, 0.2)$, which allows us to match the shape and amplitude of the non-thermal pressure contribution as parametrised by~\cite{2002.01934}, while accounting for the uncertainty in the accuracy of this parametrisation.

    The resulting constraints are shown in the \al{third} panel of Fig. \ref{fig:gamma-MC-constraints}, with the results from $C_\ell^{\gamma X}$ and $C_\ell^{\gamma X}+C_\ell^{\gamma y}$ shown in blue and green, respectively. Comparing to the results found with our minimal model (shown in the first panel of the same figure), we see that allowing for the presence of non-thermal pressure produces an upwards shift in the preferred value of $\Gamma$, and a downwards shift in $M_c$, thus bringing the joint constraints into better agreement with the values preferred by $C_\ell^{\gamma X}$. Specifically, the joint analysis of $C_\ell^{\gamma X}$ and $C_\ell^{\gamma y}$ leads to the following constraints
    \begin{align*}
      \lMc &= 14.40^{+0.12}_{-0.19},\\
      \Gamma &= 1.178^{+0.019}_{-0.016},\\
      \alpha_T &= 1.29^{+0.14}_{-0.18}.
    \end{align*}

  \subsubsection{Systematics in the Compton-$y$ maps}\label{sssec:res.params.tszsys}
    The reconstruction of the thermal SZ signal from CMB intensity maps is affected by foreground emission, dominated by Galactic and extragalactic dust, as well as radio point sources. To quantify the sensitivity of our constraints to these sources of contamination, we carried out two tests.

    First, we repeated our analysis for two alternative maps of the Compton-$y$ parameter, those released by \cite{chandran2023} and \cite{mccarthy2024} (C23 and M23 hereon, respectively). These maps use different releases of the \planck data (PR4, as opposed to our fiducial map, constructed from the official 2015 release), and were constructed using different flavours of the ``needlet ILC'' technique (NILC). The constraints in the $(\lMc,\Gamma)$ plane  found using these alternative maps are shown in Fig.~\ref{fig:gamma-MC-constraints}. Focusing on the polytropic index $\Gamma$, which is the main source of tension as shown in Section \ref{sssec:res.params.minimal}, we obtain the following constraints from the joint analysis of $C_\ell^{\gamma X}$ and $C_\ell^{\gamma y}$:
    \begin{align*}
      \Gamma &= 1.166^{+0.013}_{-0.011} \quad ({\rm C23}), \\
      \Gamma &= 1.182^{+0.013}_{-0.011} \quad ({\rm M23}),
    \end{align*}
    in comparison with our constraint $\Gamma=1.154^{+0.015}_{-0.012}$ using Planck $y$ map.
    In both cases we observe a shift towards larger values of $\Gamma$, bringing the constraints into better agreement with the $C_\ell^{\gamma X}$ data.
    
    Secondly, we repeat our analysis choosing a more conservative scale cut in the $C_\ell^{\gamma y}$ correlation. Specifically, we discard the largest angular scales, with average bin center $\ell<200$, where contamination from Galactic dust could lead to a misestimation of the power spectrum covariance, or even a bias in the measurement. As before, we observe an upwards shift in the preferred value of $\Gamma$, finding $\Gamma=1.210^{+0.014}_{-0.012}$, in better agreement with the value preferred by $C_\ell^{\gamma X}$.

    It is important to note that, in the three cases explored in this section, this shift in $\Gamma$ occurs along the same degeneracy line in the $\Gamma$-$\mathrm{log}_{10}(M_c)$ for the $C_\ell^{\gamma X}$ data that we found in our minimal analysis. It is therefore difficult to quantify whether this shift is significant or compatible with a statistical fluctuation due to variations in the low-$\ell$ power spectrum. A visual inspection of the power spectra recovered from C23 and M23 does not reveal any significant systematic shifts with respect to our fiducial measurements. We will explore the potential of systematics in the tSZ maps as a solution to the tension found within our minimal model by comparing their predictions against external data in the next section. 

    Finally, as a further test for the potential of contamination in the $y$ map from extragalactic foregrounds, correlated with the large-scale structure, we repeat our analysis using the CIB-deprojected tSZ map made available by M23. We find that this does not significantly impact our results regarding the comparison with the model constraints obtained from $C_\ell^{\gamma X}$, with a measured polytropic index $\Gamma = 1.186^{+0.015}_{-0.012}$. The main effect is limited to a vertical shift in the $M_c-\alpha_T$ degeneracy measured from $C_\ell^{\gamma y}$ and displayed in Fig.~\ref{fig:baseline_contours}. The constraints on $\tilde{\alpha}_T$ (see Eq. \ref{eq:alphat}) from the CIB-deprojected map are 
    \begin{equation}
      \tilde{\alpha}_T=0.888 \pm 0.050,
    \end{equation}
    corresponding to a $-2.2\sigma$ shift with respect to $1$. 

  \subsection{Comparison with external datasets}\label{ssec:res.extern}
    \begin{figure}[t!]
      \centering
      \includegraphics[width=\columnwidth]{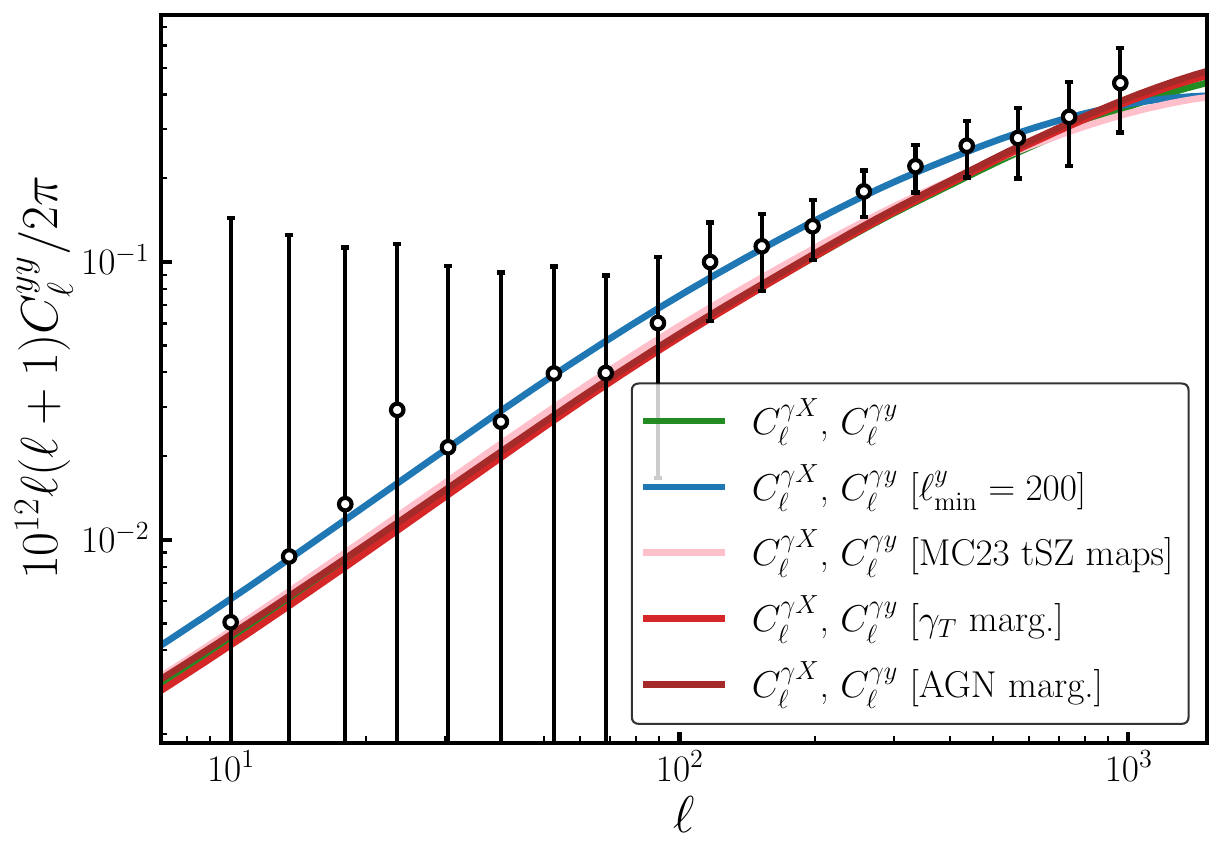}
      \caption{Comparison of theory predictions obtained from parameter constraints shown in Fig.~\ref{fig:gamma-MC-constraints} with a foreground and cosmology marginalised measurement of the tSZ auto-spectrum from~\cite{bolliet2018}. Almost all best-fit models provide similar theory predictions for this observable, which is therefore not very informative for model selection.}
      \label{fig:ext-bolliet}
    \end{figure}

    Since the combination of $C_\ell^{\gamma X}$ and $C_\ell^{\gamma y}$ allows us to constrain all the free parameters of our minimal model, we can explore the physical consistency of the derived constraints by comparing the predictions of this preferred model against external datasets that we did not use to constrain it.

    First, we use measurements of the tSZ auto-correlation, presented in \cite{bolliet2018}, and obtained from the 2015 \planck tSZ map. These measurements, marginalised over CMB foreground contamination, are shown in Figure~\ref{fig:ext-bolliet}. The figure also shows the best-fit predictions for this observable obtained from the combination of $C_\ell^{\gamma X}$ and $C_\ell^{\gamma y}$ analysed under the minimal hydrodynamic model (green), as well as including non-thermal pressure (pink), and marginalising over the AGN contamination (red). The predictions from all these models are remarkably similar, and follow the measurements reasonably well. We do not perform a quantitative goodness-of-fit analysis, since we do not have access to the full covariance matrix of these measurements. The figure also shows the best-fit model obtained using a more conservative large-scale cut for $C_\ell^{\gamma y}$ ($\ell>200$, blue), which also provides a reasonable fit to the data. In summary, the model constraints obtained under our different analysis choices all provide reasonable predictions for the tSZ auto-correlation. Although not shown in the figure, it is worth noting that the predictions obtained from the best-fit model derived from the $\gamma y$ power spectrum alone also fit the $C_\ell^{yy}$ measurements well. It is therefore not entirely surprising that all best-fit models derived from data combinations that include this cross-correlation provide reasonable predictions for $C_\ell^{yy}$.

    \begin{figure}[t!]
      \centering
      \includegraphics[width=\columnwidth]{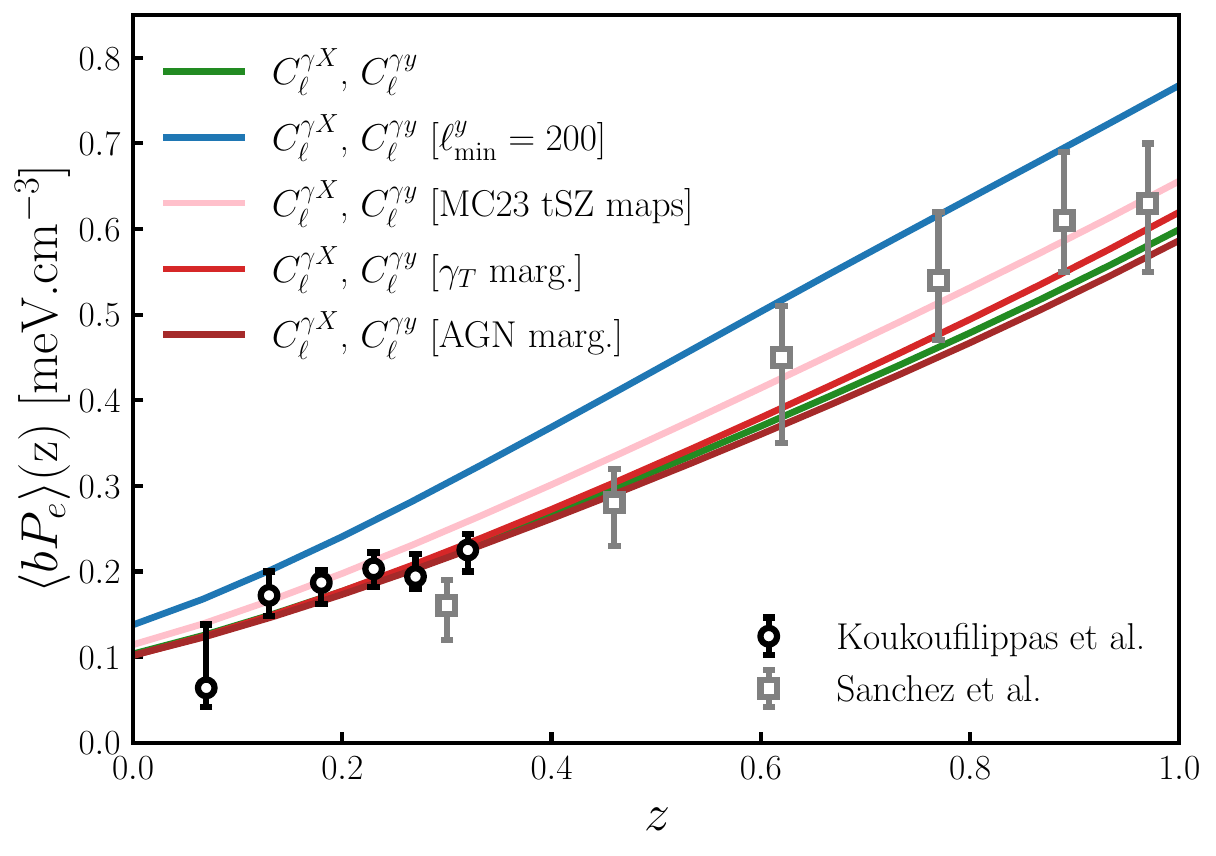}
      \caption{Average bias weighted gas pressure as a function of redshift measured in Refs~\cite{koukoufilippas2019,sanchez2023} along with theory predictions from the parameter constraints shown in Fig.~\ref{fig:gamma-MC-constraints}. All best-fit models are able to provide a reasonable description for this observable.}
      \label{fig:ext-bpe}
    \end{figure}

    Secondly, we use measurements of the bias-weighted mean pressure $\langle b P_e\rangle$, obtained from the analysis of the cross-correlation of tSZ maps with tomographic samples of galaxies. Several groups have carried out measurements of this quantity using different tSZ maps and galaxy surveys \citep{1608.04160,1904.13347,koukoufilippas2019,2006.14650,2102.07701,sanchez2023,2206.05689}. Here we use the measurements of \cite{koukoufilippas2019}, using the \planck 2015 Compton-$y$ map and galaxies from the 2MPZ and WISE$\times$SuperCOSMOS photometric surveys \citep{1311.5246,1607.01182}, and those of \cite{sanchez2023}, using tSZ maps constructed from the combination of \planck and the South Pole Telescope, and galaxies from the 3-year DES data release.  We show these measurements in Fig. \ref{fig:ext-bpe}. The theoretical prediction for $\langle bP_e\rangle$ is given by Eq. \ref{eq:bU}, with our model for the electron pressure profile replacing the generic halo quantity $U$. The figure also shows the predictions derived from the best-fit models obtained under different analysis choices, in all cases using $C_\ell^{\gamma X}$ and $C_\ell^{\gamma y}$. The prediction from the minimal hydrodynamic model seems to provide an excellent fit to the data. The predictions from the extensions including non-thermal pressure and residual AGN contamination, as well as the prediction obtained using the M23 map, are remarkably similar to the minimal model prediction. In contrast, the results using more conservative scale cuts in $C_\ell^{\gamma y}$ lie more significantly above the measurements, particularly at low redshifts. It is worth noting that this comparison is a non-trivial test of the physical consistency of our preferred models: while the auto-correlation $C_\ell^{yy}$ is dominated by the 1-halo term and is thus sensitive mostly to very massive halos at $z\sim0$, the $\langle bP_e\rangle$ measurements span a significantly larger range of redshifts and halo masses. Nevertheless, as before, we find that the predictions obtained with most of the different analysis choices are also able to reproduce these measurements with a reasonable level of accuracy (except perhaps in the case of $\ell_{\rm min}^y=200$).

    \begin{figure}[t!]
      \centering
      \includegraphics[width=\columnwidth]{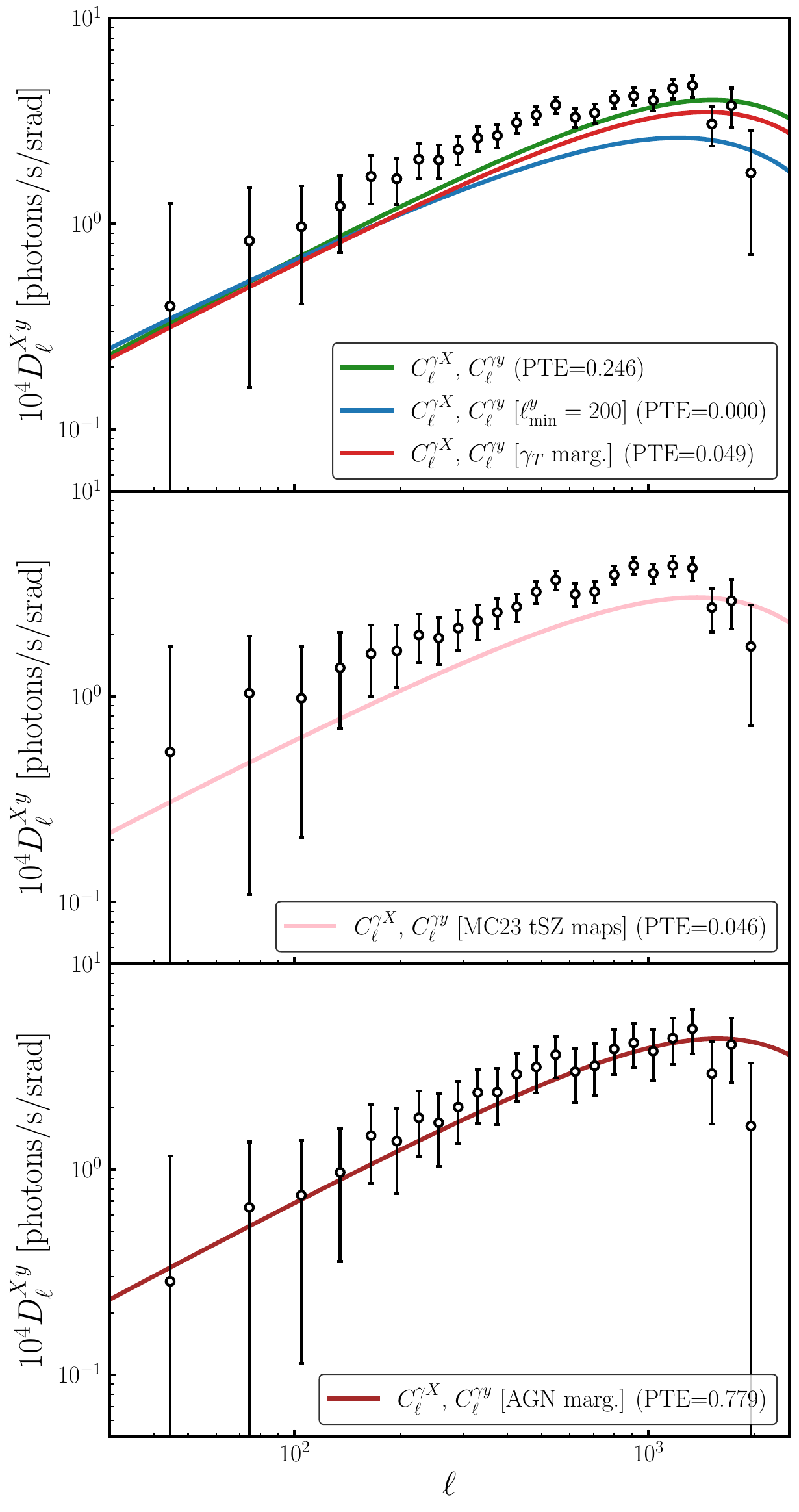}
      \caption{Angular power spectrum measurement from \xray and tSZ data. We display this cross correlation using ROSAT \xray and \planck 2015 Compton-$y$ map (top panel), using \planck PR4 MC23 maps (middle panel). We also display the correlation using ROSAT \xray and \planck 2015 marginalising over AGN contamination (bottom panel). We compare these measurements to theory predictions from the different analysis settings from Fig.~\ref{fig:gamma-MC-constraints}. Note that the covariance of this power spectrum receives a major contribution from the connected trispectrum, increasing the bin-to-bin correlations. All predictions displayed in the figure are not informed from the $C_\ell^{Xy}$ measurement. For visualisation purpose, we used $D_\ell=\ell(\ell+1)C_\ell/2\pi$.}
      \label{fig:ext-xy}
    \end{figure}

    Finally, we compare the predictions obtained from our different data combinations and analysis choices for the cross-correlation between the \xray and $y$ maps, $C_\ell^{X y}$. In the following we only focus on constraints from datasets which do not include the $Xy$ cross-correlation. We measure this power spectrum directly from the data, and estimate its covariance matrix, including its leading non-Gaussian contributions, as described in Section \ref{ssec:meth.cls}. The measurements and predictions are shown in Fig. \ref{fig:ext-xy}. Since in this case we have access to the full covariance matrix of these measurements, we can estimate the $\chi^2$ and PTE of different predictions, and these are reported in the figure. We find that the minimal model obtained from the joint fit to $C_\ell^{\gamma X}$ and $C_\ell^{\gamma y}$ is able to provide a reasonable fit to the data, with a PTE of $24.6\%$. Note that the connected 1-halo trispectrum constitutes a large contribution to the total covariance matrix of this power spectrum, since, as is the case for $C_\ell^{yy}$, it is dominated by a few large clusters at low redshifts. This means that the covariance matrix has large off-diagonal elements that can make a visual comparison between data and predictions misleading. This explains the relatively good $\chi^2$ of the $C_\ell^{\gamma X}+C_\ell^{\gamma y}$ prediction in  spite of the data showing correlated deviations with respect to it. On the other hand, we find that the prediction obtained using a more conservative scale cut in $C_\ell^{\gamma y}$ is not able to describe the data well (PTE$<0.1\%$). Allowing for a certain amount of non-thermal pressure in the modelling or using the MC23 Compton-$y$ map indeed allows for a higher polytropic index from the shear-tSZ correlation, but does provide a worse description of the $C_\ell^{Xy}$ measurement with PTE of $4.9\%$ and $4.6\%$ respectively. marginalising over AGN contamination turns out to be successful at reconciling measurements of $\Gamma$ from $C_\ell^{\gamma X}$ and $C_\ell^{\gamma y}$ and also predicts a $C_\ell^{Xy}$ prediction in excellent agreement with the data with a PTE of $77.9\%$.

\subsection{Constraints on gas thermodynamics from $\gamma y\;+\; Xy$ }

    Since we have access to the covariance of the $C_\ell^{Xy}$ measurements, and their cross-covariance with the two other cross-correlations used in our main analysis, we can also explore the ability of $C_\ell^{Xy}$ to constrain our hydrodynamic model. In particular, it is interesting to explore the possibility of replacing the $C_\ell^{\gamma X}$ data with $C_\ell^{Xy}$, combining it with $C_\ell^{\gamma y}$. The impact of AGN contamination (the most important systematic for \xray cross-correlations) should be different for these two datasets, since both tSZ and \xray maps are tracers of the hot gas, while cosmic shear maps the full matter distribution within which both gas and AGNs reside. By studying this data combinations, we will also be able to quantify whether the different model extensions explored in the previous section are fully able to alleviate the model tension between different datasets. We thus repeat our analysis using  $C_\ell^{\gamma y}+C_\ell^{Xy}$ to constrain the parameters of the hydrodynamic model. When accounting for AGN contamination, we follow the same procedure used for $C_\ell^{\gamma X}$: in the fiducial analysis we limit ourselves to masking the 2RXS sources, whereas in the case of AGN marginalisation, we estimate the contribution from resolved and unresolved AGNs theoretically and marginalise over its amplitude (the level of contamination expected in this case is shown in the right panel of Fig. \ref{fig:cls_agn}).
    
    The constraints resulting from this data combination are shown in Fig. \ref{fig:gamma-MC-constraints} (red contours). We see that, in our minimal analysis, the constraints obtained overlap with those from the $C_\ell^{\gamma X}+C_\ell^{\gamma y}$, although they exhibit the same level of tension with the parameter values favoured by $C_\ell^{\gamma X}$. More interestingly, the variations of our analysis that explore the impact of systematics in the tSZ data (using more conservative scale cuts for $y$ and employing the M23 tSZ map) yield parameter constraints when using $C_\ell^{\gamma y}+C_\ell^{Xy}$ that still show tension with the constraints from $C_\ell^{\gamma X}$. The presence of tSZ systematics on its own is therefore not a satisfactory solution to the model discrepancy observed in the minimal analysis. The inclusion of non-thermal pressure in the model leads to constraints from $C_\ell^{\gamma y}+C_\ell^{Xy}$ that are in somewhat better agreement with those found for $C_\ell^{\gamma X}$ and $C_\ell^{\gamma X}+C_\ell^{\gamma y}$, although the tension between different datasets does not appear to be completely resolved. Finally, the \al{second} panel of this figure shows that the marginalisation over residual AGN contamination leads to model constraints that agree well with the regions of parameter space favoured by the other data combinations explored in this analysis.

    In summary, the comparison of our results against external data seems to suggest that the most likely avenues to explain the internal tensions apparent in our minimal hydrodynamic models are extending this model to include two additional astrophysical effects: non-thermal pressure support and contribution from unresolved AGNs. Both of these effects are known to be present in the observables studied here at some level, and they allow us to describe all of them in a self-consistent manner.

\section{Conclusion}\label{sec:conc}
  \begin{figure}[t!]
    \centering
    \includegraphics[width=\columnwidth]{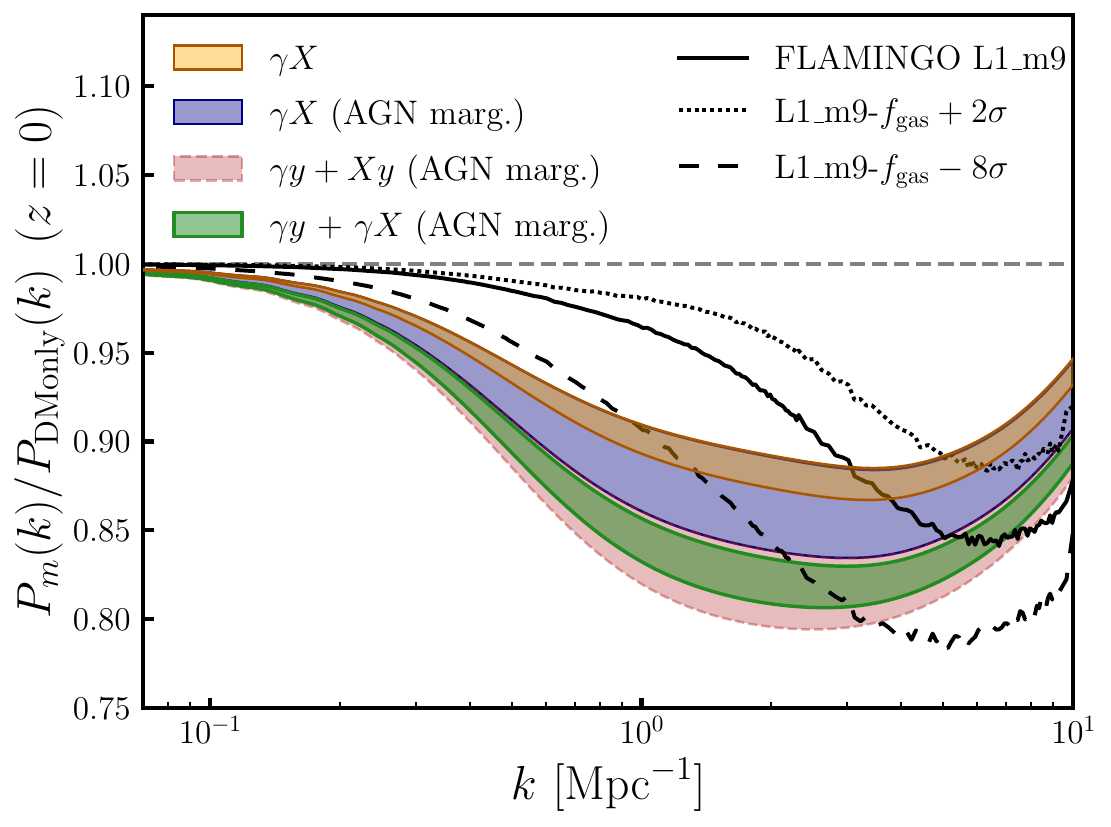}
    \caption{68\% constraints on the matter power spectrum suppression factor derived from the AGN marginalised constraints combining the $\gamma \times X$ and $\gamma \times y$ correlations (green). We also display constraints on the power suppression from $\gamma \times X$ alone marginalising over AGN contamination (blue) or not accounting for it (orange), as well as the combination of shear-tSZ correlations and the $X\times y$ power spectrum (red). We compare these constraints to predictions from the FLAMINGO simulations~\cite{2306.04024,2410.17109} where $f_\mathrm{gas}-8\sigma$ and $f_\mathrm{gas}+2\sigma$ corresponds to stronger and weaker baryonic feedback respectively.}
    \label{fig:baryon-suppression}
  \end{figure}

  We have presented a joint analysis of the cross-correlations between cosmic shear data (from DES-Y3) and two probes of the hot gas in the intergalactic medium: maps of the \xray intensity (from ROSAT), and of the Compton-$y$ parameter (from \planck). As probes of gas, these datasets are sensitive to complementary properties of the distribution, chemical composition, and thermodynamics of the gas. Their combined analysis within a single, self-consistent physical model should, in principle, break the main parameter degeneracies present in individual probes, and provide a more complete picture of the physics of the cosmic baryon component.

  We have explored a particularly simple model for the hot gas, depending on three free parameters. These are: the halfway mass $\lMc$, which controls the fractional amount of gas bound in dark matter haloes, the polytropic index $\Gamma$, which determines the concentration and scale dependence of this gas, and $\alpha_T$, which parametrises deviations of the gas temperature from perfect virialisation. We have also extended this minimal model in two different ways: we have accounted for the presence of non-thermal pressure support (vital in connecting the density and temperature of the gas), and included a model for the contribution to the \xray intensity from unresolved non-thermal point sources, primarily AGNs.

  We find that this minimal hydrodynamic model is able to describe both cross-correlations well, and that the corresponding best-fitting parameters are also able to predict a number of external datasets that were not used in the analysis (the tSZ auto-correlation, tomographic measurements ot the bias-weighted mean gas pressure, and the \xray-tSZ cross-correlation). However, when studying the the regions of parameter space preferred by different data combinations, we find tensions between $C_\ell^{\gamma X}$ and $C_\ell^{\gamma y}$, particularly in terms of $\Gamma$ (even if this tension does not prevent the joint best-fit from describing the data well). We have explored various ways to alleviate this tension, including the two extended models mentioned above, as well as variations in the choices made in the analysis of the Compton-$y$ map. Based on their ability to predict the external datasets, and in particular to provide consistent constraints when including the tSZ-\xray cross-correlation (see \al{second} panel of Fig. \ref{fig:gamma-MC-constraints}), we find that the most promising solutions to this tension are marginalising over residual AGN contamination and accounting for non-thermal pressure.
  
  Although both of these effects are sub-dominant, affecting the signal at the level of a few tens of percents, current cross-correlation measurements are sensitive enough that this can have a significant impact on the model constraints. Future cross-correlation analyses involving high-SNR gas probes will therefore need to rely on more sophisticated models than the relatively simple one explored here. This will be particularly important for any studies aiming to analyse these probes in combination with cosmic shear auto-correlations to deliver constraints on cosmological parameters marginalised over a self-calibrated model for baryons \cite{2109.04458,2404.06098}. In this sense, the model presented here can be improved along several angles:
  \begin{itemize}
    \item It will be important to study the impact of metallicity gradients on the predicted \xray signal  \cite{astro-ph/0012232}. A significant deviation of the density-squared-weighted average metallicity from the value assumed here (and a potential dependence on halo properties) would affect the amplitude and the small-scale dependence of the \xray cross-correlations and their interpretation in terms of gas properties.
    \item The model should include a more flexible parametrisation for the small-scale clustering of AGNs, beyond the simple bias scaling used here, and ideally calibrated against simulations. This would allow for a more robust marginalisation over uncertainties in the AGN contribution.
    \item Our measurements are sensitive to scales straddling the regimes dominated by the 1-halo and 2-halo contributions to the matter power spectrum. Although this transition occurs on significantly smaller scales for tSZ (and likely also \xray) cross-correlations \cite{2005.00009}, our model will benefit from including corrections to the usual under-estimation of power in this transition regime within the halo model. This could be achieved using a response approach, as in \cite{2005.00009}, the prescription of \cite{2011.08858}, emulator-based methods \cite{1811.09504}, or recent advances to describe halo non-linear bias \cite{1910.07097,2101.12187}.
  \end{itemize}
  We envisage including these improvements, and quantifying the accuracy of the resulting model predictions against hydrodynamical simulations in future work.

  The constraints found on our hydrodynamic model can be used to make predictions regarding the impact of baryonic effects in cosmological weak lensing analyses. In short, the main impact of baryonic physics in the small-scale matter power spectrum is the ejection of gas from dark matter haloes due to AGN-driven outflows. The ejected gas, at significantly lower density and temperature than the bound component, effectively ceases to contribute to the matter clustering signal, which results in a suppression in the matter power spectrum on small scales ($0.3\,{\rm Mpc}^{-1}\lesssim k\lesssim10\,{\rm Mpc}^{-1}$) \cite{1905.06082,2305.09710,2410.17109}. The key parameter in this case is $\lMc$, which determines the halo mass range over which feedback effects dominate. To quantify the level of suppression predicted by our constraints, we follow a procedure similar to that used by HMCode~\cite{2009.01858}: we calculate the halo model prediction for the matter power spectrum, scaling, in the 1-halo term, the matter density profile by the factor $f_{\bf CDM}+f_{\bf b}(M)$ (thus quantifying the impact of gas mass loss due to feedback), and including the contribution from the central stellar component. The baryonic suppression factor $S(k)\equiv P_m(k)/P_{\rm DMO}(k)$ is then calculated as the ratio of the power spectrum computed using the bound fraction $f_{\bf b}(M)$ given by our measurement of $\lMc$, and the one given by $f_{\bf b}+f_{\bf CDM}=1$ with no stellar contribution. We verified that the suppression factor thus calculated is robust against the detailed modelling of the 1-halo to 2-halo transition regime. 
  
  The result of this calculation is shown in Fig. \ref{fig:baryon-suppression}. The figure shows the suppression predicted by our constraints on $\lMc$ using $C_\ell^{\gamma X}$ ignoring the contribution from unresolved AGNs, and using $C_\ell^{\gamma X}$ and $C_\ell^{\gamma X}+C_\ell^{\gamma y}$ marginalising over this contribution. The main impact of AGN contamination is allowing for lower values of $\lMc$ to compensate for the fraction of the signal due to AGNs, leading to a larger ejected fraction and therefore higher baryonic suppression factors. The uncertainties in the predicted suppression factor grow significantly after marginalising over AGNs, although a similar level of precision is then recovered when including $C_\ell^{\gamma y}$. For comparison, the figure also shows the baryonic suppression recovered by the FLAMINGO hydrodynamical simulations for different levels of feedback intensity \cite{2410.17109}. The level of suppression recovered by our constraints lies closer to that predicted by the simulations with stronger AGN feedback. This is in approximate qualitative agreement with the findings of \cite{2404.06098,2410.19905} based on kSZ observations, although the level of suppression found in \cite{2404.06098} on small scales ($k> 1\,h{\rm Mpc}^{-1}$) appears to be significantly larger than our predictions. The Figure also shows the constraints found from the combination of $C_\ell^{\gamma y}$ and $C_\ell^{X y}$ (i.e. ignoring the shear-\xray cross-correlation). Interestingly, the constraining power in this case is comparable to that of $C_\ell^{\gamma X}+C_\ell^{\gamma y}$. Thus, the tSZ-\xray cross-correlation can serve as a useful alternative to $C_\ell^{\gamma X}$ in constraining baryonic effects, given the different impact of AGN contamination on both cross-correlations. Finally, we must also note the qualitative difference in the shapes of the baryonic suppression factors predicted here and found in simulations, likely a consequence of the relatively simple HMCode-like prescription used here. A thorough validation of the relatively simple hydrodynamical model used here is therefore necessary before quantitative conclusions can be drawn from our constraints regarding the impact of baryonic effects in the context of the ongoing $S_8$ tension\al{, particularly in the context of recent results highlighting that other sources of systematic uncertainty (e.g. photometric redshift calibration) could be behind it \cite{2503.1944}.}

\section*{Acknowldedgements}
  We thank Raul Angulo, Giovanni Aric\'o, Ian McCarthy, Will McDonald, Jaime Salcido, Tilman Tr\"oster, and Matteo Zennaro for useful comments and discussions. ALP and DA acknowledge support from a Science and Technology Facilities Council Consolidated Grant (ST/W000903/1). TF is supported by a Royal Society Newton International Fellowship. DA and CGG acknowledge support from the Beecroft Trust.  We made extensive use of computational resources at the University of Oxford Department of Physics, funded by the John Fell Oxford University Press Research Fund. The power spectrum measurements used in this analysis are available at \url{https://github.com/Cosmotheka/Cosmotheka_likelihoods/tree/main/papers/syx}~\cite{cls_data}.

\emph{Software}:  We made extensive use of the {\tt numpy} \citep{oliphant2006guide, van2011numpy}, {\tt scipy} \citep{2020SciPy-NMeth}, {\tt astropy} \citep{1307.6212, 1801.02634}, {\tt healpy} \citep{Zonca2019}, {\tt matplotlib} \citep{Hunter:2007} and {\tt GetDist} \citep{Lewis:2019xzd} python packages, as well as the \hpx package \cite{astro-ph/0409513}.

This paper makes use of software developed for the Large Synoptic Survey Telescope. We thank the LSST Project for making their code available as free software at \url{http://dm.lsst.org}. 

\emph{Data:} 

  \emph{DES}. This project used public archival data from the Dark Energy Survey (DES). Funding for the DES Projects has been provided by the U.S. Department of Energy, the U.S. National Science Foundation, the Ministry of Science and Education of Spain, the Science and Technology Facilities Council of the United Kingdom, the Higher Education Funding Council for England, the National Center for Supercomputing Applications at the University of Illinois at Urbana-Champaign, the Kavli Institute of Cosmological Physics at the University of Chicago, the Center for Cosmology and Astro-Particle Physics at the Ohio State University, the Mitchell Institute for Fundamental Physics and Astronomy at Texas A\&M University, Financiadora de Estudos e Projetos, Funda{\c c}{\~a}o Carlos Chagas Filho de Amparo {\`a} Pesquisa do Estado do Rio de Janeiro, Conselho Nacional de Desenvolvimento Cient{\'i}fico e Tecnol{\'o}gico and the Minist{\'e}rio da Ci{\^e}ncia, Tecnologia e Inova{\c c}{\~a}o, the Deutsche Forschungsgemeinschaft, and the Collaborating Institutions in the Dark Energy Survey.
        
  The Collaborating Institutions are Argonne National Laboratory, the University of California at Santa Cruz, the University of Cambridge, Centro de Investigaciones Energ{\'e}ticas, Medioambientales y Tecnol{\'o}gicas-Madrid, the University of Chicago, University College London, the DES-Brazil Consortium, the University of Edinburgh, the Eidgen{\"o}ssische Technische Hochschule (ETH) Z{\"u}rich,  Fermi National Accelerator Laboratory, the University of Illinois at Urbana-Champaign, the Institut de Ci{\`e}ncies de l'Espai (IEEC/CSIC), the Institut de F{\'i}sica d'Altes Energies, Lawrence Berkeley National Laboratory, the Ludwig-Maximilians Universit{\"a}t M{\"u}nchen and the associated Excellence Cluster Universe, the University of Michigan, the National Optical Astronomy Observatory, the University of Nottingham, The Ohio State University, the OzDES Membership Consortium, the University of Pennsylvania, the University of Portsmouth, SLAC National Accelerator Laboratory, Stanford University, the University of Sussex, and Texas A\&M University.
    
  Based in part on observations at Cerro Tololo Inter-American Observatory, National Optical Astronomy Observatory, which is operated by the Association of Universities for Research in Astronomy (AURA) under a cooperative agreement with the National Science Foundation.

  For the purpose of open access, a CC BY public copyright license is applied to any Author Accepted Manuscript version arising from this submission.

\onecolumngrid
\newpage

\appendix
\section{AGN contribution}\label{app:agn}
  \begin{figure}
    \centering
    \includegraphics[width=0.49\textwidth]{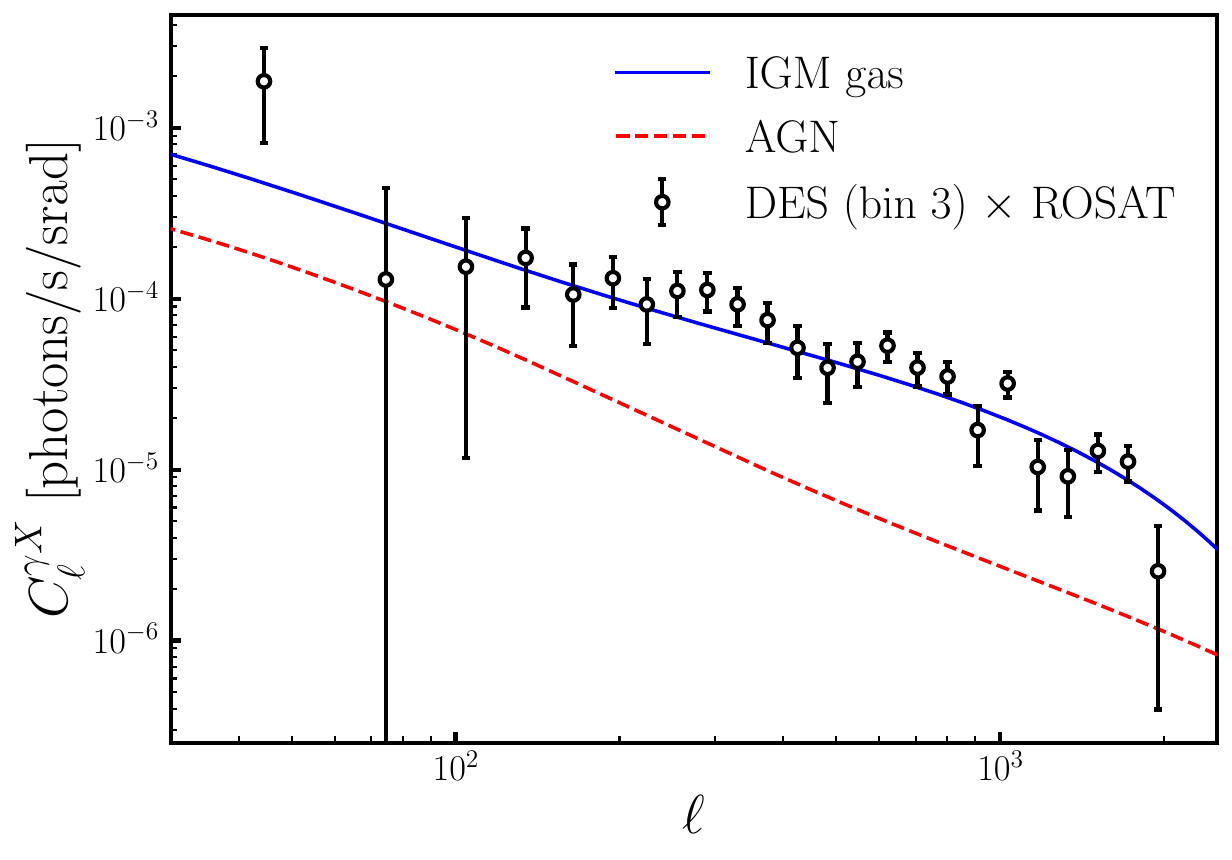}
    \includegraphics[width=0.49\textwidth]{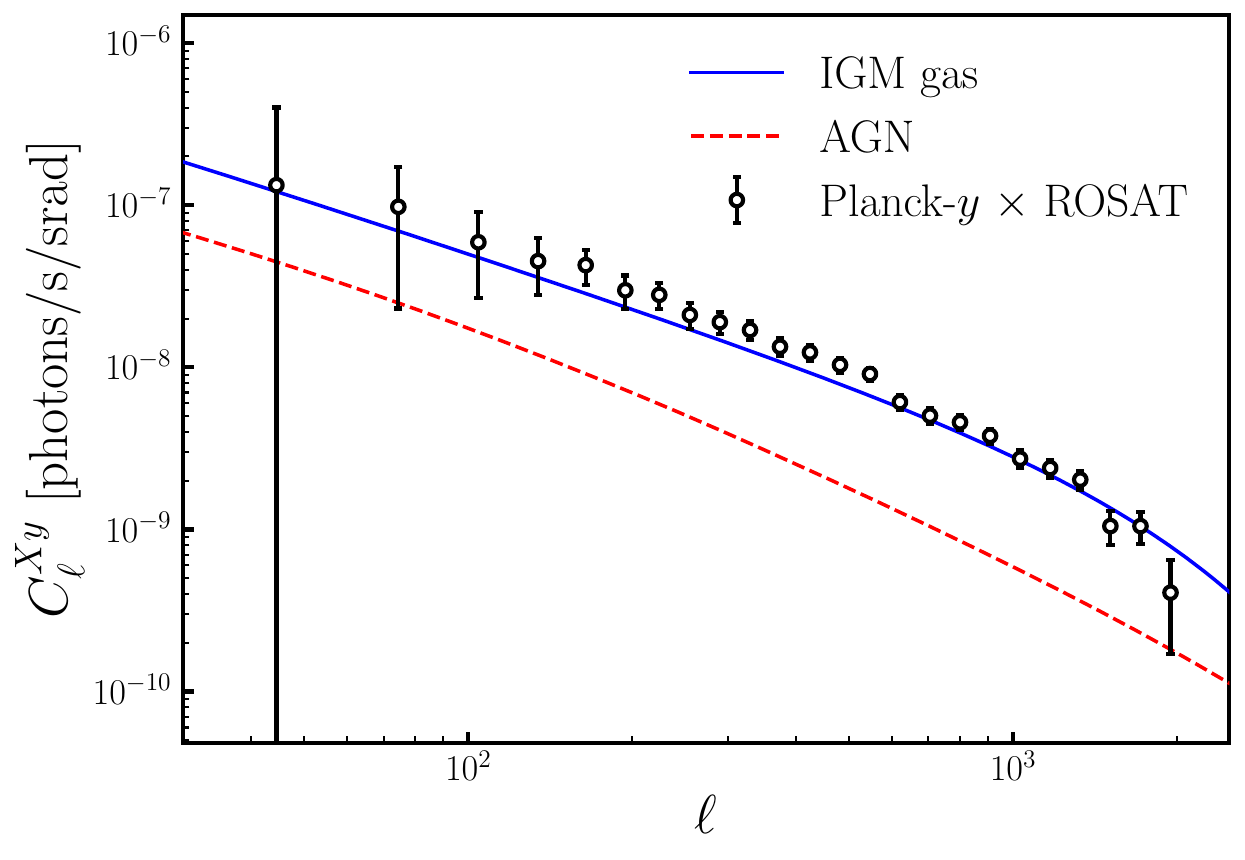}
    \caption{{\sl Left:} cross-correlation between cosmic shear in the third DES redshift bin and \xrays from ROSAT. The figure shows our measurements (black dots with error bars), together with the predicted contribution from IGM gas (blue) and AGN (dashed red). AGN contribute to $\sim20\%$ of the signal within the model described in the text. {\sl Right:} same results for the tSZ-\xray cross-correlation.}
    \label{fig:cls_agn}
  \end{figure}
  Consider a sample of \xray sources with a luminosity function $dn/d\log_{10}L$, representing the comoving number density of sources within a logarithmic luminosity interval, where the luminosity $L$ is defined in a particular rest-frame energy band $\phi_{\rm ref}(\energ)$:
  \begin{equation}
    L\equiv\int\,d\energe\,\phi_{\rm ref}(\energe)\,\frac{dE_e}{dt_e\,d\energe},
  \end{equation}
  where $dE_e/dt_ed\energe$ is the energy spectrum of a given source. For simplicity, we will assume that sources of a given type (e.g. absorbed or unabsorbed AGN) have spectra with the same energy dependence $dE_e/dt_e\,d\energe\propto f(\energe)$. The emissivity (i.e. number of photons emitted per unit time, volume, and energy interval) is therefore
  \begin{equation}
    \left.j_\energ\right|_{\rm AGN}=(1+z)^3\int d\log_{10}L\,\frac{dn}{d\log_{10}L}\,\frac{1}{\energ_e}\frac{dE_e}{dt_e\,d\energe},
  \end{equation}
  where the prefactor $(1+z)^3$ accounts for the fact that $dn/d\log_{10}L$ is a comoving density.
    
  Substituting this in Eq. \ref{eq:crgen}, we obtain:
  \begin{equation}
    \crate_{\rm AGN}(\nv)=\int\frac{d\chi}{4\pi}\left\langle\frac{A}{\energ}\right\rangle_z\,\rho_L(\chi\nv,z),
  \end{equation}
  where we have defined the band-averaged detector area over energy
  \begin{equation}\label{eq:mean_Aoe}
    \left\langle\frac{A}{\energ}\right\rangle_z\equiv\frac{\int d\energ_o\,\phi(\energo)\,A(\energo)\,\frac{f(\energo(1+z))}{\energo(1+z)}}{\int d\energ\,\phi_{\rm ref}(\energ)\,f(\energ)},
  \end{equation}
  and the luminositity density $\rho_L$
  \begin{equation}\label{eq:rhoL}
    \rho_L\equiv\int d\log_{10}L\,\frac{dn}{d\log_{10}L}\,L.
  \end{equation}
  Note the different bandpasses used in the numerator and denominator of Eq. \ref{eq:mean_Aoe}, corresponding to the instrumental bandpass within which observations are made, $\phi$, and the one over which luminosities entering the luminosity function are defined $\phi_{\rm ref}$. Separating the luminosity function into its background mean value and the spatial fluctuations due to inhomogeneities in the distribution of AGN, with overdensity $\delta_s$, we obtain Eq. \ref{eq:crAGN}.

  We use the luminosity function measurements and parametrisation of \cite{1503.01120}. Specifically, we use the flexible double-power law (FDPL) parametrisations for absorbed and unabsorbed AGN. Note that, in this case, the luminosity function is defined in the hard \xray band ($\energ\in[2\,{\rm keV},10\,{\rm keV}]$), and thus AGN spectra must be redshifted to the soft ROSAT band used here. We assume a power-law spectrum for unabsorbed AGN, with a spectral index $\Gamma_\energ=-1.9$, and an absorbed spectrum that decreases by a factor $10^{-2}$ at $\energe=2\,{\rm keV}$ (see Fig. 4 in \cite{1503.01120}). Finally, we must adopt a model for the clustering of AGN. Using the results of \cite{2301.01388}, we assume a linear biasing relation $\delta_s=b_{\rm AGN}\delta_M$, with $b_{\rm AGN}=1$.

  Figure \ref{fig:cls_agn} shows the AGN contribution to our measurements of $C_\ell^{\gamma X}$ and $C_\ell^{Xy}$ from AGN, together with our data and the predicted contribution from hot gas. Within this model, AGN contamination is of the order of 15-20\% of the signal. As discussed in the main text, the relative simplicity of the model used, and existing uncertainties regarding the clustering properties of unresolved \xray sources imply that the real level of contamination could be substantially larger, depending on the redshifts and angular scales explored.

\section{Joint constraints from \texorpdfstring{$C_\ell^{\gamma X}$, $C_\ell^{\gamma y}$ and $C_\ell^{Xy}$}{gamma-X, gamma-y and Xy power spectra}}\label{app:full_combo}

\begin{figure}[h!]
    \centering
    \includegraphics[width=0.7\textwidth]{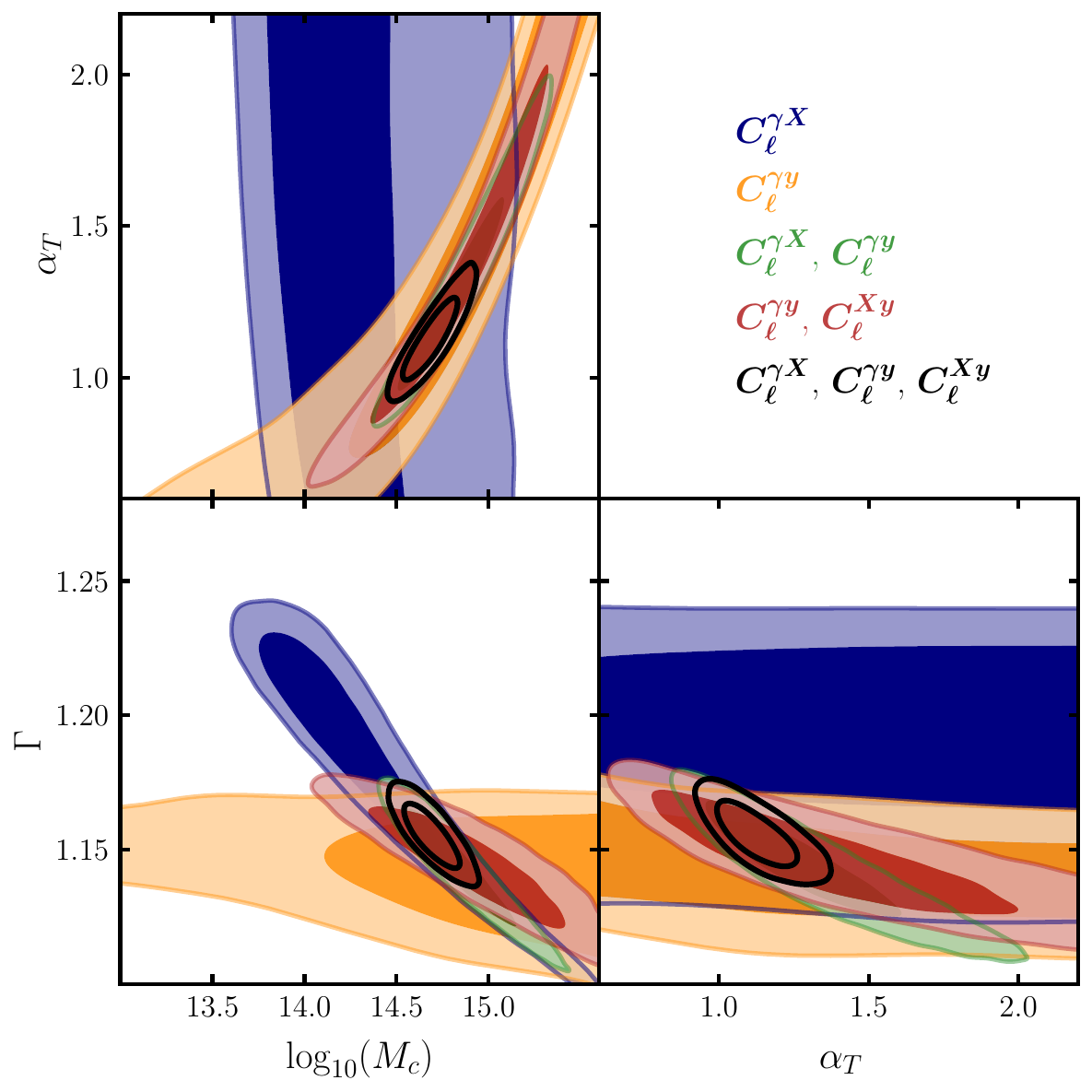}
    \caption{2D marginalised posterior distributions for the halo model parameters $\lMc$, $\Gamma$ and $\alpha_T$ after marginalisation over AGN contamination in the \xray signal. This figure shows the constraints derived from the full power spectrum data vector, including the $\gamma X$, $\gamma y$ and $Xy$ correlations.}
    \label{fig:full_combo}
  \end{figure}

  \al{As pointed out in Section~\ref{sssec:res.params.agn}, after marginalization over AGN contamination, $C_\ell^{\gamma X}$, $C_\ell^{\gamma y}$ and $C_\ell^{Xy}$ measurements provide consistent constraints on the hydrodynamical halo model studied in this paper. In Figure~\ref{fig:full_combo}, we present joint constraints from all three correlations (black solid line), reducing parameter posterior errors with respect to our baseline data combination ($C_\ell^{\gamma y}$, $C_\ell^{\gamma X}$). The most significant impact is on the parameter $\alpha_T$, with $\alpha_T=1.13^{+0.08}_{-0.10}$, indicating no departure from the virial relation.}

\twocolumngrid

\bibliography{draft}

\end{document}